\documentclass[12pt]{article}
\usepackage{mathrsfs}
\usepackage{natbib,graphicx,subfigure,setspace,lscape,longtable}
\usepackage{natbib,epsfig,graphicx,bm,bbm}
\usepackage{mathrsfs,amsmath,amsthm,amssymb,color}

\usepackage{indentfirst}
\usepackage{hyperref}
\hypersetup{
        citecolor=black,
	colorlinks=true,
	linkcolor=black,
	filecolor=black,
	urlcolor=black,
}

\bibpunct{(}{)}{;}{a}{,}{,}

\newtheorem{theorem}{Theorem}
\newtheorem{lemma}{Lemma}

\newtheorem{proposition}{Proposition}

\usepackage{array}
\newcolumntype{H}{@{}>{\lrbox0}l<{\endlrbox}}

\def\beq{\begin{equation}}
\def\eeq{\end{equation}}
\def\beqr{\begin{eqnarray}}
\def\eeqr{\end{eqnarray}}
\def\beqrs{\begin{eqnarray*}}
\def\eeqrs{\end{eqnarray*}}
\def\bet{\begin{theorem}}
\def\eet{\end{theorem}}
\def\bel{\begin{lemma}}
\def\eel{\end{lemma}}
\def\bep{\begin{proposition}}
\def\eep{\end{proposition}}
\def\bg{\begin{figure}[tbph]\begin{center}}
\def\eg{\end{center}\end{figure}}

\def\bc{\begin{center}}
\def\ec{\end{center}}

\def\n{\nonumber}

\def\beq{\begin{equation}}
\def\eeq{\end{equation}}
\def\beqr{\begin{eqnarray}}
\def\eeqr{\end{eqnarray}}
\def\beqrs{\begin{eqnarray*}}
\def\eeqrs{\end{eqnarray*}}
\def\bet{\begin{theorem}}
\def\eet{\end{theorem}}
\def\bel{\begin{lemma}}
\def\eel{\end{lemma}}
\def\bep{\begin{proposition}}
\def\eep{\end{proposition}}
\def\bg{\begin{figure}[tbph]\begin{center}}
\def\eg{\end{center}\end{figure}}

\def\bc{\begin{center}}
\def\ec{\end{center}}

\def\n{\nonumber}

\def\wt{\widetilde}
\def\wh{\widehat}

\def\mB{\mathcal B}

\def\mR{\mathbb{R}}

\def\mS{\mathcal S}

\def\mR{\mathbb{R}}
\def\mS{\mathcal S}

\def\var{\mbox{var}}
\def\cov{\mbox{cov}}

\numberwithin{equation}{section}

\usepackage{array}
\newcolumntype{H}{>{\setbox0=\hbox\bgroup}c<{\egroup}@{}}

\newcommand{\Cov}{\textnormal{Cov}}

\newcommand{\ve}{{\varepsilon}}
\renewcommand{\epsilon}{{\ve}}
\renewcommand{\hat}{\widehat}
\def\wt{\widetilde}
\renewcommand{\tilde}{\wt}

\usepackage[margin=1in]{geometry}

\numberwithin{equation}{section}

\title{\bf{Optimal Subsampling Bootstrap for Massive Data}}
\author{Yingying Ma\footnote{School of Economics and Management, Beihang University, China. Email: mayingying@buaa.edu.cn.} \quad\quad Chenlei Leng\footnote{Department of Statistics, University of Warwick, UK. Email: C.Leng@warwick.ac.uk.} \quad\quad Hansheng Wang\footnote{Guanghua School of Management, Peking University, China. Email: hansheng@gsm.pku.edu.cn.}}
\date{}
\begin{document}

\maketitle
\begin{abstract}
The bootstrap  is  a widely used procedure for  statistical inference because of  its simplicity and attractive statistical properties. However, the vanilla version of bootstrap is no longer feasible computationally for many  modern massive datasets due to the need to repeatedly resample the entire  data. Therefore, several improvements to the bootstrap method have been made in recent years, which assess the quality of estimators by subsampling the full dataset before resampling the subsamples.
  Naturally, the performance of these modern subsampling methods is influenced by tuning parameters such as the size of subsamples,   the number of subsamples, and the number of resamples per subsample. In this paper,  we develop  a novel hyperparameter selection methodology for selecting these tuning parameters. Formulated as an optimization problem to find the optimal value of some measure of accuracy of an estimator subject to computational cost, our framework provides closed-form solutions for the optimal hyperparameter values for subsampled bootstrap, subsampled double bootstrap and bag of little bootstraps, at no or little extra time cost. Using the mean square errors as a proxy of the accuracy measure, we apply our methodology to study, compare and improve the performance of these modern versions of bootstrap developed for massive data through numerical study. The results are promising.

\end{abstract}

\section{Introduction}
Real data analysis often faces situations where statistical inference is not tractable. This can happen if the interest is to estimate the variance of an estimator that is not easily estimable or when a robust estimator of the variance is warranted under the suspicion that assumptions are invalid. In this case, the bootstrap method \citep{efron1990more,efron1994introduction} provides a simple elegant solution for automatic inference, enabling the computation of various inference quantities without the need to know their analytical formula.
Under certain conditions, many bootstrap estimators  are  generally consistent \citep{van1996weak} and   can  be  more accurate than those based on asymptotic approximation \citep{hall1994methodology}.

Traditional bootstrap (TB) methods were first developed for small datasets for which computation was not an issue. In particular, the vanilla version of the bootstrap estimates parameters of interest by repeatedly resampling $N$ observations with replacement from $N$ data points in the original sample and thus is feasible only if the sample size is relatively modest and the computation is performed on a single computer \citep{booth1994monte}.
With the emergence of Big Data, these methods are no longer applicable to modern datasets that are massive in size. First and foremost, it is no longer the case that these datasets can be loaded into the main memory.  One approach to mitigate this problem is to use parallel and distributed computing systems to divide a dataset, estimate the parameters of interest on each computer, and aggregate these estimates on a central machine. This approach usually involves high communication costs between different computer nodes and, thus, is sometimes not desirable
\citep{Runzeli2013}, although progress has been made to alleviate this \citep{jordan2019communication,volgushev2019distributed,Chen2019,
Chen2021,fan2021communication}.
The main focus of this paper is to develop methods for optimal bootstrap inference on a single computer which will free us from the concern on communication cost.

 Recognizing the limitation of TB for big datasets,  subsampling methods that only requires repeated computation of the estimator for subsamples with size much smaller than the original dataset,  have been developed \citep{Politis1999}. A leading example is the so called $n$-out of-$N, $ bootstrap \citep{bickel1997resampling}, in which each subsample draws $n$ observations out of $N$ points often with $n \ll N$. We will refer to this bootstrap scheme as subsampling bootstrap and abbreviate it as SB for short hereafter. Although SB reduces the size of each subsample from $N$ in TB to $n$, \cite{bickel2008on} showed that its performance is rather sensitive to the choice of  subsample size. Moreover, SB must perform a rescaling of their output which requires knowledge and explicit use of the convergence rate of the estimator, making it  less automatic to deploy than TB. The above limitations of SB prompted \cite{kleiner2014scalable} to introduce  the method of Bag of Little Bootstraps
(BLB). Similar to SB, BLB starts by subsampling the whole data with the size of each subsampled subset much smaller than $N$ and then follows up by resampling the subsets via simple random sampling with replacement. Crucially, the resampling in the second step of BLB is  performed in such a way that the size of each resample is $N$, the same as the size of the entire data. The computational saving of BLB over that of TB roots in the fact that the maximum number of distinct elements in each resample is bounded by the size of each subsample in the first step. For many estimators such as M-estimators obtained  via empirical risk minimization, this means that we just need to optimize a weighted loss function with a smaller number of distinct items in the empirical risk than that of TB. Compared to SB, BLB  requires no analytical re-scalling because of the size of the resamples in step two and thus makes a fully automatic  method for statistical inference
\citep{kleiner2014scalable}.

Intuitively, the performance  of BLB  relies  on three hyperparameters
\begin{itemize}
\item $n$ (the size of the bootstrap subsamples or subsets in step one),
\item $R$ (the number of the subsamples or replicates in step one),
\item $B$ (the total number of resamples per subset in step two).
\end{itemize} With unlimited computational resources, those  hyperparameters should be  set as large as possible for reliable inference.
With limited computational budget, however, these three  hyperparameters need to  be selected. Indeed, it is found that the performance of BLB can be sensitive to the choice of these parameters as discussed in  \cite{kleiner2014scalable} and \cite{sengupta2016subsampled}. Because of this, \cite{sengupta2016subsampled} raises an open question:
\begin{center}
\textit{How do we optimally choose $n$, $B$ and $R$ in BLB to balance statistical accuracy and running time?}
\end{center}
In \cite{kleiner2014scalable}, they suggested an adaptive method for selecting $R$ and $B$. By introducing a new tolerance parameter, they tracked resamples for each subset until that tolerance level was reached. Without knowing the variability of the precision estimate though, it is unclear how this additional hyperparameter should be specified for a given computational cost.
As a partial solution, \cite{sengupta2016subsampled} proposed an alternative approach named subsampled double bootstrap (SDB), which simply sets $B=1$ as in BLB. This certainly solves the problem of choosing between $R$ and $B$. However, the relationship between the subset size $n$ and the number of replicates $R$ remains elusive. In order to choose $n$ and $R$ in SDB, an intuitive approach is to choose $n$ as large as possible as long as $n$ data points can fit on a computer while adjusting the size of $R$ such that the computational budget is not exceeded. As we show however, the choice of $n$ and $R$ does impact the performance of SDB.

In this paper,  we  develop a novel framework to find  the optimal balance between statistical efficiency and computational cost for these modern variants of the bootstrap method. To strike this balance, an intuitive procedure is to maximize a certain measure of this efficiency subject to a running time constraint. Intuitively, any measure of efficiency will depend on the hyperparameters $n$, $R$, and $B$ and so will the running time. Denoting these dependencies as $\mbox{Efficiency}(n, R, B)$ and $\mbox{Time}(n, R, B)$ respectively, our general framework seeks to identify  $(n^*,  R^*, B^*)$ such that
\[ (n^*, R^*, B^*) = \arg\mbox{optim}_{n, R, B}~~ \mbox{Efficiency}(n, R, B), ~~\mbox{subject to}~ \mbox{Time}(n, R, B) \le C_{\max},
 \]
 where $C_{\max}$ is a given computational cost allocated to bootstrap and $\arg\mbox{optim}$ reads as the argument that optimizes the subsequent function.  For the constraint $C_{\max}$, it is often more convenient to specify it corresponding to the time needed to implement a bootstrap method with some prespecified $(n, R, B)$ combination. In this case, we can immediately understand that our framework aims to improve the performance of BLB with prespecified $n, R$ and $B$, by using an optimal combination of these hyperparameters obtained via optimization.

Often for mathematical convenience, it is easier to work with the inefficiency of an estimator based on various bootstrap implementations, for example by examining some loss function or the variance of an estimator.  Intuitively the dependence of the inefficiency on these hyperparameters can be characterized by  terms usually inversely proportional to $n$, $R$ and $B$ or simple functions of them. This is because conditional on data, bootstrap subsamples and resamples can be viewed as independent and identically distributed \citep{efron1990more,efron1994introduction}. The measure of inefficiency of interest often depends on the empirical distribution of these bootstrap subsamples and resamples. This is in sharp contrast to  $n$-out of-$N$ bootstrap where explicit knowledge of the convergence rate must be known for it to be applicable.

To illustrate our general framework, we use the mean square error (MSE) of an estimator as a proxy for the accuracy measure and discuss the use of various bootstrap methods for estimating the standard errors of the sample mean. The sample mean estimates the population mean which is a quantity of major interest in many statistical problems \citep{he1996a}.
 By a careful theoretical analysis, we find that the asymptotic efficiency of various bootstrap estimates are closely related to the hyperparameters.  In particular, these relationships for BLB and SDB are analytically simple, allowing us to identify the optimal combinations of the hyperparameters in these procedures in closed-form. Thus, we go beyond providing an affirmative answer to the open question raised in \cite{sengupta2016subsampled} by presenting closed-form solutions to the optimal hyperparameter triple $(n, R, B)$ for BLB. Furthermore, our procedure provides the optimal pair of parameters in $n$ and $R$ for SB in  \cite{sengupta2016subsampled}. Our approach can be readily extended to deal with multivariate random variables and general parameters and we discuss how this can be done.   Although we only consider the independent data, by a similar approach, this method can be extended to block bootstrap designed for dependent data.
  We show via extensive simulations that our approach improve the performance of BLB and SB with similar computational budget. Thus  our answer to the open question posed in  \cite{sengupta2016subsampled} is confirmed empirically.   Note that an early approach to find the optimal  resampling size in the double bootstrap appeared in
    \cite{booth1994monte} but  their method  is  computationally infeasible  for massive data.

   The rest of the article is organized as follows. Section 2 introduces different bootstrap methods and their associated theoretical properties. Our hyperparameter selection approach is presented in  Section \ref{sec: optimal}. We discuss the estimation of more general parameters and statistics  in Section \ref{sec:general}.
   Extensive numerical study is conducted in Section 4. The article is concluded with a discussion in Section 5. All the theoretical conditions,  proofs, and additional numerical results are found in the Supplementary Material.

The following notations are used throughout the paper. For two positive sequences $\{a_n \}$, $\{ b_n\}$, we use $a_n \lesssim b_n$ to mean $a_n \le C b_n$ for some constant $C$ independent of $n$, $a_n \gtrsim b_n$ to mean $b_n \lesssim a_n$, and $a_n\asymp b_n$ to mean $a_n \lesssim b_n$ and $b_n \lesssim a_n$. When stating any results, we always assume   that the sample size satisfies $N\to\infty$. We use $U(a, b)$ to denote a uniform random variable on $[a,b]$. Finally, for a matrix
 $A=(a_{ij})\in \mathbb{R}^{m\times n}$, we denote $\|A\|_F=(\sum_{i,j}a_{ij}^2)^{1/2}$ as its Frobenius norm.

\section{Bootstrap Methods for Univariate Random Variable }
For better illustration of our approach, we start by examining the simplest case where the statistic of interest is the mean of a univariate random variable. The theory to be presented hereafter is further  developed for multivariate  random variables   in Section \ref{sec:multi} and we discuss general statistics in Section \ref{sec:general}.
 To  facilitate the theoretical development, we need the
following technical conditions.
\begin{itemize}
\item [(C1)](TB Condition)  As $N \rightarrow \infty$, we assume $B\rightarrow \infty$.
\item [(C2)] (BLB  Condition) As $N \rightarrow \infty$, we assume $\min\{n,R,B\}\rightarrow \infty$.
In addition assume that $n<N$, $N=o(nBR)$ and $n\asymp R \asymp B$.
\item [(C3)] (SDB and SB  Condition) As $N \rightarrow \infty$, assume $\min\{n, R\} \rightarrow \infty$.
In addition assume that $n<N$ and $R  \gtrsim  n$.
 \end{itemize}

The above conditions are mild and reasonable. By assuming these relationships between different hyperparameters and the whole sample size $N$,  they essentially require that  $n, R,$ and  $B$ should be large enough  to facilitate an asymptotic analysis of higher
order terms.

Let $\mS=\{X_1, \cdots,  X_N\}$ be the sample where $X_1, X_2,\cdots,$ $X_N \in \mR^1$  are independent  and identically  distributed random variables with mean  $\mu$ and variance  $\sigma^2$. We assume that the centered $j$ th moment of $X_i$ exists such that $E(X_i-\mu)^j =\sigma_j < \infty$ for $3\leq j\leq 6$.  A simple estimator of the mean parameter $\mu$ is the same average denoted as
$\overline X =N^{-1}\sum_{i=1}^{N}X_i.$
The estimation accuracy of $\overline X$ is measured by its standard error (SE) which equals $\sigma/\sqrt{N}$ analytically. This SE can be consistently estimated as
\begin{equation}
\hat{\mbox{SE}^2} = \hat\sigma^2/N, \label{eq:fulldataestimator}
\end{equation}
where $\hat \sigma^2= \sum_{i=1}^{N} (X_i-\overline X)^2/N$.
We characterize the mean and the variance of $\hat{\mbox{SE}^2}$  in the following theorem.

\begin{theorem} For the estimator in \eqref{eq:fulldataestimator},
we  have
\beqr
E(\hat{\mbox{SE}}^2)=\frac{\sigma^2}{N}\Big(1-\frac{1}{N}\Big)
~\mbox{and}~
\var(\hat{\mbox{SE}}^2)= \frac{\sigma_4 - \sigma^4}{N^3}\Big \{1 +o(1) \Big\}. \n
\eeqr
\end{theorem}
\noindent
By this theorem,   the MSE of $\hat{\mbox{SE}}^2$ is dominated by its variance and can be seen as
\beqr
\mbox{MSE}(\hat{\mbox{SE}}^2) =\frac{\sigma_4-\sigma^4}{N^3}\Big \{1 +o(1)\Big\},
\label{MSE_a}
\eeqr
which scales  inverse-proportionally to $N^3$ in the leading order.

\subsection{ Traditional bootstrap} \label{sec:bt}
We analyze the MSE of the traditional bootstrap or the vanilla bootstrap estimator of SE in this subsection.
Denote $B$ as the total number of bootstrap resamples that are sampled uniformly with replacement from the original sample. That is, for any  $b=1,\cdots, B$,
 the $b$th bootstrap sample is obtained as  $\mB^{(b)}=\{X_{1}^{(b)}, X_{2}^{(b)}, \cdots, X_{N}^{(b)} \}$, where
 $X_{i}^{(b)}$ is independently generated via simple random sampling with replacement from the whole sample $\mS$.
From $\mB^{(b)}$, the $b$th bootstrap sample mean can be calculated  as
$\overline X^{(b)}= N^{-1}\sum_{i=1}^{N} X_{i}^{(b)}$. Subsequently, an estimate of SE$^2$  via the bootstrap method is seen as
\beqr
\hat {\mbox{SE}}^2_{\tiny\mbox{TB}}  = B^{-1} \sum_{b=1}^{B} \Big(\overline X^{(b)}- \overline X \Big)^2. \n
\eeqr
Conditional on $\mS$, for any $b$,
 $X^{(b)}_{i}$'s  are independent and identically distributed due to the bootstrap scheme. This immediately implies that
 $P(X_{i}^{(b)}=X_j)=1/N$ for any $1 \leq j\leq N$. That is, each element in $\mS$ has equal probability of being sampled.
Accordingly, we have $E(\overline X^{(b)}|\mS)= E(X_{i}^{(b)}|\mS)=\overline X$
and $\var(\overline X^{(b)}|\mS)= N^{-1}\var(X_{i}^{(b)}|\mS)=N^{-1} \hat\sigma^2.$
We have the following theorem for this traditional bootstrap estimator of $\mbox{SE}^2$.
\begin{theorem}
		For the traditional bootstrap estimator,  assume Condition (C1) holds,
	 we have
\beqr
E(\hat{\mbox{SE}}^2_{\tiny\mbox{TB}}) =\frac{\sigma^2}{N }\Big(1-\frac{1}{N}\Big) ~~\mbox{and}~~
\var(\hat{\mbox{SE}}^2_{\tiny\mbox{TB}}) =\var (\hat{\mbox{SE}}^2)\Big(\frac{2\sigma^4}{\sigma_4- \sigma^4}\cdot \frac{N}{B} +1 \Big)\Big \{1+o(1) \Big\}.\n
\eeqr
\end{theorem}
\noindent
From this theorem, we can immediately obtain  the MSE of $\hat{\mbox{SE}}^2_{\tiny\mbox{TB}}$ as
 \beqr
 \mbox{MSE}(\hat{\mbox{SE}}^2_{\tiny\mbox{TB}}) &=& \var (\hat{\mbox{SE}}^2)\Big(\frac{2\sigma^4}{\sigma_4- \sigma^4}\cdot \frac{N}{B} +1 \Big)\Big \{1+o(1) \Big\} + \frac{\sigma^4}{N^4}\n\\
 &=& \Big( \frac{\sigma_4-\sigma^4}{N^3} + \frac{2 \sigma^4  }{N^2 B} \Big)\Big\{1 + o(1)\Big\}.
 \label{MSE_b}
 \eeqr
This suggests  that $B$, the number of bootstrap resamples,
needs to be   the same order of $N$ or larger, for
$\mbox{MSE}(\hat{\mbox{SE}}^2_{\tiny\mbox{TB}})$  to achieve the same convergence  rate as
$\mbox{MSE}(\hat{\mbox{SE}}^2)$ in \eqref{MSE_a}. Given a computational budget, the most efficient traditional bootstrap estimator is to keep drawing resamples until this budget runs out.

 \subsection{ Bag of little bootstraps}
 We next provide a brief review of the Bag of Little Bootstraps (BLB)  method in   \cite{kleiner2014scalable}, which is carried out via a two-step procedure including a subsampling and a resampling step.
 \vspace{-0.3cm}
 \begin{itemize}
\item[Step 1.] The subsampling step: We draw $R$ subsamples or little bootstrap samples, each of which, denoted as $\mS^{(r)}=\{X^{(r)}_{1},X^{(r)}_{2},\cdots, X^{(r)}_{n}\}$ for $r=1,\cdots, R$, is drawn via simple random sampling  with  replacement from the whole sample $\mS$.
Note that the size of $\mS^{(r)}$ is $n$, which is usually much smaller than $N$.
\item[Step 2.] The resampling step: $B$ weighted resamples are drawn such that the cardinality of each resample is $N$. Specifically, denote the $b$th resample as  $\mB^{(r,b)}= \{  X_{1}^{(r,b)}, X_{2}^{(r,b)}, \cdots, X_{N}^{(r,b)}\}$. For each $r$ and $b$,  $X_{i}^{(r,b)}$ is independently drawn via simple random sampling from $\mS^{(r)}$ with replacement. This scheme implies that the number of distinct points in each bootstrap resample $\mB^{(r,b)}$ is $n$ at maximum.
\end{itemize}

From $\mB^{(r,b)}$, the target statistic $\overline X$ can be estimated by the sample average in the resample as
\beq
\overline X^{(r,b)}= N^{-1} \sum_{i=1}^N   X_{i}^{(r,b)} =  N^{-1}  \sum_{i=1}^n  X^{(r)}_{i} f_{i}^{(r,b)}, \n
\eeq
 where  $ f_{i}^{(r,b)}$ is the number of times that $X_{i}^{(r)}$ appears in  $\mS^{(r)}$.
 That is
 $ f_{i}^{(r,b)}=\sum_{j=1}^N  I(X_{i}^{(r,b)}= X_{j}^{(r)})$ where $I(\cdot)$ is the indicator function.
 Obviously, the random vector $ f^{(r,b)}=(f_{1}^{(r,b)},$ $ f_{2}^{(r,b)}, \cdots, f_{n}^{(r,b)})^\top\in \mR^n$ follows a multinomial distribution with parameter $N$ and $p$ with $p=( 1/n,1/n,\cdots, 1/n)^\top\in \mR^n$.

As the result of the two-step procedure, each bootstrap resample $\mB^{(r,b)}$ has cardinality $N$, but with at most $n$ distinct elements. Equivalently, each $\mB^{(r,b)}$ can be seen as a weighted resample of size $n$. Thus, the BLB  avoids the need for repeated computation on resamples having size comparable with that of the original data set, since each BLB resample contains at most $n$ distinct elements.  In a large class of estimators commonly encountered, including M-estimators, computation can take weighted data representation. It is for these estimators that BLB has huge computational advantages. In comparison to the traditional bootstrap,  the storage requirement for the BLB method is substantially less demanding as long as $n \ll N$, and for these estimators,
the cost of computing the estimator based on the BLB resamples will be substantially lower than that based on the traditional bootstrap resamples.

 Given the bootstrap resamples $\overline X^{(r,b)}$,  SE$^2$ can be estimated as
 \beqr
\hat {\mbox{SE}}^2_{\tiny\mbox{BLB}}= \frac{1}{RB} \sum_{r=1}^R   \sum_{b=1}^{ B} \Big(\overline X^{(r,b)}- \overline X^{(r)}\Big)^2, \label{SE_c} \n
\eeqr
where $\overline X^{(r)} =n^{-1}\sum_{i=1}^{n}X^{(r)}_{i}$.

We now derive the MSE of $\hat {\mbox{SE}}^2_{\tiny\mbox{BLB}}$. Note that  conditional on $\mS$ and $\mS^{(r)}$,
 $\overline X^{(r,b)}$'s  are independent and identically distributed.
Accordingly, we have $E(\overline X^{(r,b)}|\mS,\mS^{(r)})=  \overline X^{(r)}$
and $\var(\overline X^{(r,b)}|\mS,\mS^{(r)})=  N^{-1} \wt\sigma_{r}^2,$
 where $\wt \sigma_{r}^2= n^{-1}\sum_{i=1}^{n} ( X^{(r)}_{i}-\overline X^{(r)})^2$.
 We have the following theorem.
  \begin{theorem}
For the BLB estimator, assume Condition (C2) holds, we have
\beqr
   E(\hat {\mbox{SE}}^2_{\tiny\mbox{BLB}})
   &=&  \frac{ \sigma^2}{N} (1-\frac{1}{n}) \Big\{1+ o(1)\Big\},
   \n\\
\var (\hat{\mbox{SE}}^2_{\tiny\mbox{BLB}})  &=&  \var (\hat{\mbox{SE}}^2)\Big(  \frac{2\sigma^4}{\sigma_4- \sigma^4}\cdot\frac{ N} {RB}+\frac{N}{nR}+1\Big)\Big\{1+ o(1)\Big\}.\n
   \eeqr
\end{theorem}
\noindent
From the above theorem, the MSE of $\hat{\mbox{SE}}^2_{\tiny\mbox{BLB}}$ is immediately seen as
 \beqr
\mbox{MSE}(\hat{\mbox{SE}}^2_{\tiny\mbox{BLB}})& =& \var (\hat{\mbox{SE}}^2)\Big(  \frac{2\sigma^4}{\sigma_4- \sigma^4} \cdot \frac{ N} {RB}+\frac{N}{nR}+1\Big)\Big\{1+ o(1)\Big\} +\frac{\sigma^4}{N^2n^2}\n\\
&=&\Big(  \frac{\sigma_4-\sigma^4}{N^3} + \frac{2\sigma^4}{N^2RB}+ \frac{\sigma_4-\sigma^4}{N^2nR}+\frac{\sigma^4}{N^2n^2} \Big) \Big\{1+ o(1)\Big\}.
\label{MSE_c}
\eeqr
Compared with \eqref{MSE_a}, for the MSE of $\hat{\mbox{SE}}^2_{\tiny\mbox{BLB}}$ to be comparable with that of the analytical one, we require $n \gtrsim \sqrt{N}$, $nR \gtrsim N$ and $RB \gtrsim N$ which hold by taking $n \asymp \sqrt{N}$, $R \asymp \sqrt{N}$ and $B\gtrsim \sqrt{N}$ for example. Alternatively, we can take $n \sim N^{\omega}$ with $\omega \in[0.5,1]$, $R \sim N^{1-\omega}$ and $B \sim N^{\gamma}$ with $\gamma>0.5$.
\subsection{ Subsampled   bootstrap}

The BLB method  is closely related to the so-called ``$n$-out of-$N$'' bootstrap method or subsampled bootstrap studied in
\cite{bickel1997resampling}. With some abuse of notation, let $R$ be the total number of subsamples. For any $1\leq r \leq R$, we use $\mS^{(r)}=\{X_{i}^{(r)}: 1\leq i\leq n\}$ to denote the $r$th subsample, where
$X_{i}^{(r)}$ is generated  independently  by  simple random sampling with replacement from the whole sample $\mS$. Note that the number  of distinct elements in each subsample is at most $n$.  Based on $\mS^{(r)}$, the target statistic  $\overline X$
can be  computed as
$\overline X^{(r)}= n^{-1} \sum_{i=1}^n   X_{i}^{(r)}. $
Accordingly,   SE$^2$  can be estimated by
\beqr
\hat {\mbox{SE}}^2_{\tiny\mbox{SB}}= \Big(\frac{n}{N}\Big){ R}^{-1} \sum_{r=1}^{R} \Big(\overline X^{(r)}-\overline X\Big)^2. \label{SE_d} \n
\eeqr
Comparing    $\hat {\mbox{SE}}^2_{\tiny\mbox{SB}}$ with $\hat {\mbox{SE}}^2_{\tiny\mbox{TB}}$, we find that a re-scaling factor  $n/N$ is  needed  for $\hat {\mbox{SE}}^2_{\tiny\mbox{SB}}$ which  requires the knowledge of the convergence rate of the target estimator.  This makes the SB
 less automatic \citep{kleiner2014scalable,sengupta2016subsampled}.

We now study  the MSE of $\hat{\mbox{SE}}^2_{\tiny\mbox{SB}}$. Note that
 conditional on $\mS$, for any given $r$,
 $X_{i}^{(r)}$s  are independent and identically distributed with
 $P(X_{i}^{(r)}=X_j)=1/N$ for any $1 \leq j\leq N$.
Accordingly, we have $E(\overline X^{(r)}|\mS)= E(X_{i}^{(r)}|\mS)=\overline X$
and $\var(\overline X^{(r)}|\mS)= n^{-1}\var(X_{i}^{(r)}|\mS)=n^{-1} \hat\sigma^2.$
We have the following theorem.
  \begin{theorem}
For the SB estimator, assume Condition (C3) holds,
	 we  have
\beqr
E(\hat {\mbox{SE}}^2_{\tiny\mbox{SB}})
   &=&N^{-1} (1-n^{-1} )\sigma^2, \n \\
\var (\hat{\mbox{SE}}^2_{\tiny\mbox{SB}}) &=&\var (\hat{\mbox{SE}}^2)\Big(\frac{2\sigma^4}{\sigma_4-\sigma^4}\cdot\frac{N}{R}+1\Big) \Big\{1+o(1)\Big\}. \n
   \eeqr
\end{theorem}
\noindent
Thus,  the MSE for $\hat{\mbox{SE}}^2_{\tiny\mbox{SB}}$ can be expressed as
 \beqr
 \mbox{MSE}(\hat{\mbox{SE}}^2_{\tiny\mbox{SB}}) &=& \var (\hat{\mbox{SE}}^2)\Big(\frac{2\sigma^4}{\sigma_4-\sigma^4}\cdot\frac{N}{R}+1\Big)\Big \{1+o(1)\Big\}  +
 \frac{ \sigma^4}{N^2n^2} \n\\
 &=& \Big( \frac{ \sigma_4-\sigma^4}{N^3} +\frac{2\sigma^4}{N^2R}+\frac{\sigma^4}{N^2n^2}\Big) \Big \{1+o(1)\Big\}.
  \label{MSE_d}
 \eeqr
It is seen that for the MSE of the subsampled bootstrap to be comparable to that of the analytical estimator in \eqref{MSE_a}, we require $n \gtrsim \sqrt{N}$ and $R\gtrsim N$.

\subsection{ Subsampled double bootstrap}

Motivated by the  BLB method and the subsampled bootstrap, \cite{sengupta2016subsampled} proposed the so-called  subsampled double bootstrap (SDB).  The SDB takes a similar two-step approach as in the BLB, with the crucial difference of taking a single resample in the resampling step of BLB. Specifically,
in the first step,  SDB randomly draw  $R$ subsamples of the data, denoted as  $\mS^{(r)}=\{X^{(r)}_{1},X^{(r)}_{2},\cdots, X^{(r)}_{n}\}$,
 where  $X^{(r)}_{i}$'s are independently
 generated by  simple random sampling with replacement from the whole sample $\mS$.
 In the second stage, we only generate one bootstrap resample  from each subset $\mS^{(r)}$, which is denoted as $\mB^{(r,1)}=\{ X^{(r,1)}_{1}, X^{(r,1)}_{2},\cdots, X^{(r,1)}_{N}\}$.
 Here $X^{(r,1)}_{i}$'s are  independently  generated from  $\mS^{(r)}$ by simple  random sampling with replacement.
 Based on $\mB^{(r,1)}$,  $\overline X$ can be  estimated  as
$
\overline X^{(r,1)}= \frac{1}{N} \sum_{i=1}^N   X_{i}^{(r,1)},  \n
$
and  SE$^2$ can be estimated by
$\hat {\mbox{SE}}^2_{\tiny\mbox{SDB}}= R^{-1} \sum_{r=1}^R    \Big(\overline X^{(r,1)}-\overline X^{(r)}\Big)^2 $
with $\overline X^{(r)} =n^{-1}\sum_{i=1}^{n}X^{(r)}_{i}$.

Conditional on $\mS$,
 $\overline X^{(r,1)}$s  are independent and identically distributed for $ 1\leq r\leq R$.
Accordingly, we have $E(\overline X^{(r,1)}|\mS,\mS^{(r)})=  \overline X^{(r)}$
and $\var(\overline X^{(r,1)}|\mS,\mS^{(r)})=  N^{-1}\wt\sigma_{r}^2,$   where $\wt \sigma_{r}^2= n^{-1}\sum_{i=1}^{n} ( X^{(r)}_{i}-\overline X^{(r)})^2$.
 We have the following theorem.
  \begin{theorem}
		For the SDB estimator, assume Condition (C3) holds,
	 we  have
\beqr
   E(\hat {\mbox{SE}}^2_{\tiny\mbox{SDB}})
   &=&  \Big(\frac{ \sigma^2}{N}- \frac{ \sigma^2}{Nn}\Big) \Big\{1+ o(1)\Big\},
   \n\\
\var (\hat{\mbox{SE}}^2_{\tiny\mbox{SDB}}) &=& \var (\hat{\mbox{SE}}^2)\Big(  \frac{2\sigma^4}{\sigma_4-\sigma^4}\cdot\frac{ N} {R}+1\Big)\Big\{1+ o(1)\Big\}.\n
   \eeqr
\end{theorem}
\noindent
From the above theorem, we can immediately obtain  the MSE  of $\hat{\mbox{SE}}^2_{\tiny\mbox{SDB}}$, which can be expressed as
\beqr
 \mbox{MSE}(\hat{\mbox{SE}}^2_{\tiny\mbox{SDB}}) &=&\var (\hat{\mbox{SE}}^2)\Big(  \frac{2\sigma^4}{\sigma_4-\sigma^4}\cdot\frac{ N} {R}+1\Big)\Big\{1+ o(1)\Big\}+
 \frac{ \sigma^4}{N^2n^2}\Big\{1+o(1)\Big\}\n\\
&=&\Big(\frac{\sigma_4-\sigma^4}{N^3} +\frac{2\sigma^4}{N^2R}+\frac{\sigma^4}{N^2n^2}\Big) \Big\{1+ o(1)\Big\}.
\label{MSE_e} \eeqr
From the above expression, for the MSE of $\hat{\mbox{SE}}^2_{\tiny\mbox{SDB}} $ to be comparable with that of the analytical estimator in \eqref{MSE_a}, we will require that
$n\gtrsim \sqrt{N}$ and $R\gtrsim N$.

\subsection{Bootstrap for multivariate  random variables }\label{sec:multi}
In this subsection, we  extend the univariate random variable to the multivariate case.
With some abuse of notations, we still use    $\mS=\{X_1, \cdots,  X_N\}$ to denote  the whole sample set where   $X_i=(X_{i,1},\cdots,X_{i,p})^\top\in \mR^p$ for each $1\leq i\leq p$,  is independent  and identically  distributed with mean  $\mu=(\mu_1,\cdots,\mu_p)^\top$ and variance  $\Sigma=(\sigma_{j,k})$. We will assume that $p\geq 2$ is fixed for easy exposition. By definition,
we have  $E(X_{i,j}-\mu_j)(X_{i,k}-\mu_k)= \sigma_{j,k}$ and  $E(X_{i,j}-\mu_j)^2= \sigma_{j,j}$. We assume that the centered fourth moment of $X_{i,j}$ exists; that is,  $E(X_{i,j}-\mu_j)^4=\sigma_{j,j,2}< \infty$ and $E(X_{i,j}-\mu_j)^2(X_{i,k}-\mu_k)^2= \sigma_{j,k,2}< \infty$ for each $1\leq j\neq k\leq p$.
The estimation accuracy of $\overline X=(\overline X_1,\cdots,\overline X_p)^\top\in\mR^p$ can be measured by $\Sigma/N$, which  can be consistently estimated via its sample analogue.

Using the vector notation for $X$, an estimate of  $\Sigma/N$ via BLB is denoted  as $\hat \Sigma_{\tiny\mbox{BLB}}$ where
\[\hat \Sigma_{\tiny\mbox{BLB}}= (RB)^{-1}\sum_{r=1}^R   \sum_{b=1}^{ B} (\overline X^{(r,b)}-\overline X^{(r)})(\overline X^{(r,b)}-\overline X^{(r)})^\top\] with its $(j,k)$th ($j\not= k$) element denoted by
$\hat{\mbox{SE}}^2_{j,k,\tiny\mbox{BLB}}$ and the $j$th diagonal term denoted as $\hat{\mbox{SE}}^2_{j,\tiny\mbox{BLB}}$.
We now derive the MSE of $\hat{\mbox{SE}}^2_{j,k,\tiny\mbox{BLB}}$ in the following theorem.
 \begin{theorem}
For the BLB estimator,
assume Condition (C2) holds, we have  for $ j\not= k $
\beqr
    \mbox{MSE}\Big(\hat{\mbox{SE}}^2_{j,k,\tiny\mbox{BLB}}\Big) = \Big(\frac{ \sigma_{j,k,2}+\sigma_{j,k}^2}{N^3}+\frac{\sigma_{j,j}\sigma_{k,k}+\sigma_{j,k}^2} {N^2RB}+\frac{\sigma_{j,k,2}+\sigma_{j,k}^2}{N^2nR}+\frac{ \sigma^2_{j,k}}{N^2n^2}\Big)\Big\{1+ o(1)\Big\}.\label{multi_BLB1}
   \eeqr
\end{theorem}
On the other hand, by the result of Theorem 3, we immediately have the following results for the diagonal terms
 \beqr
 \mbox{MSE}(\hat{\mbox{SE}}^2_{j,\tiny\mbox{BLB}})=\Big(  \frac{\sigma_{j,j,2}-\sigma_{j,j}^2}{N^3} + \frac{2\sigma_{j,j}^2}{N^2RB}+ \frac{\sigma_{j,j,2}-\sigma_{j,j}^2}{N^2nR}+\frac{\sigma_{j,j}^2}{N^2n^2} \Big) \Big\{1+ o(1)\Big\}.  \label{multi_BLB2}
\eeqr
Define the MSE of  $\hat \Sigma_{\tiny\mbox{BLB}}$ as the sum of the MSEs of estimating each individual element.
Combining  \eqref{multi_BLB1} and \eqref{multi_BLB2},   we further obtain  the MSE
of   $\hat \Sigma_{\tiny\mbox{SB}}$ as
\beqr
\mbox{MSE}(\hat{\mbox{SE}}^2_{\tiny\mbox{BLB}})=\Big(  \frac{c_1}{N^2RB}+ \frac{c_2}{N^2nR}+\frac{c_3}{N^2n^2} +\frac{c_2}{N^3}   \Big) \Big\{1+ o(1)\Big\}, \label{MSE_BLB}
\eeqr
where
\begin{eqnarray}
c_1&=&2\sum_{j=1}^{p}\sigma_{j,j}^2 +\sum_{j_1\neq j_2}\sigma_{j_1,j_1}\sigma_{j_2,j_2}+\sum_{j_1\neq j_2}\sigma_{j_1,j_2}^2,\label{c1}  \\
c_2&=&\sum_{j=1}^{p}(\sigma_{j,j,2}-\sigma_{j,j}^2) +\sum_{j_1\neq j_2}\sigma_{j_1,j_2,2}+\sum_{j_1\neq j_2}\sigma_{j_1,j_2}^2,  \label{c2}  \\
c_3&=&\sum_{j=1}^{p}\sigma_{j,j}^2  +\sum_{j_1\neq j_2}\sigma_{j_1,j_2}^2.\label{c3}
   \end{eqnarray}
   Note that in order to implement our optimal hyperparameter selection method,  $c_1, c_2$ and $c_3$ need to be consistently estimated.
   Since they only need to be computed once, we know that the computational cost of obtaining them is negligible
in comparison to the bootstrap procedure.

   Next, the  estimate of  $\Sigma/N$ via SB can be seen as
   \[\hat \Sigma_{\tiny\mbox{SB}}=n(N R)^{-1} \sum_{r=1}^{R} (\overline X^{(r)}-\overline X)(\overline X^{(r)}-\overline X)^\top,\] for which we have the following theorem.
    \begin{theorem}
For the SB estimator, assume Condition (C3) holds,
	 we  have
\beqr
   \mbox{MSE} (\hat{\mbox{SE}}_{j,k,\tiny\mbox{SB}})= \Big\{\frac{\sigma_{j,k}^2}{N^2n^2}+ \frac{\sigma_{j,k}^2+\sigma_{j,j}\sigma_{k,k}}{N^2R}  +\frac{\sigma_{j,k,2}+ \sigma_{j,k}^2}{N^3}\Big\}\Big\{1+o(1)\Big\},  \quad 1\leq j,k\leq p. \n
   \eeqr
   \end{theorem}
   \noindent
  Moreover,  for its diagonal terms, from Theorem 4, we can obtain
   $$\mbox{MSE}(\hat{\mbox{SE}}^2_{j,\tiny\mbox{SB}})
 = \Big( \frac{2\sigma_{j,j}^2}{N^2R}+\frac{\sigma_{j,j}^2}{N^2n^2}+\frac{ \sigma_{j,j,2}-\sigma_{j,j}^2}{N^3} \Big) \Big \{1+o(1)\Big\}, \quad 1 \leq j\leq p.$$
   Combining the above results, we further obtain  the MSE
for all the elements in  $\hat \Sigma_{\tiny\mbox{SB}}$ as
   \beqr
\mbox{MSE}(\hat{\mbox{SE}}^2_{\tiny\mbox{SB}})=\Big(  \frac{c_1}{N^2R}+ \frac{c_3}{N^2n^2} +\frac{c_2}{N^3}   \Big) \Big\{1+ o(1)\Big\}.\label{MSE_SB}
\eeqr

Lastly, we derive the  estimate of  $\Sigma/N$ via SDB denoted   as
\[\hat \Sigma_{\tiny\mbox{SDB}}=R^{-1} \sum_{r=1}^{R} (\overline X^{(r,1)}-\overline X^{(r)})(\overline X^{(r,1)}-\overline X^{(r)})^\top,\]
with its MSE characterized by the following theorem.
\begin{theorem}
For the SDB estimator, assume Condition (C3) holds,
	 we  have for $ j\not= k $
\beqr
   \mbox{MSE}(\hat {\mbox{SE}}^2_{j,k,\tiny\mbox{SDB}})
  = \Big(\frac{ \sigma_{j,k,2} + \sigma_{j,k}^2}{N^3}+3\frac{\sigma_{j,k,2}-\sigma_{j,j}\sigma_{k,k}}{N^2nR}+\frac{\sigma_{j,j}\sigma_{k,k}+\sigma_{j,k}^2} {N^2R}+\frac{ \sigma^2_{j,k}}{N^2n^2}\Big)\Big\{1+ o(1)\Big\}.\n
   \n
   \eeqr
\end{theorem}
For the diagonal terms, from Theorem 5,  we can obtain
 $$\mbox{MSE}(\hat{\mbox{SE}}^2_{j,\tiny\mbox{SDB}}) =\Big( \frac{2\sigma_{j,j}^2}{N^2R}+\frac{\sigma_{j,j}^2}{N^2n^2}+\frac{\sigma_{j,j,2}-\sigma_{j,j}^2}{N^3} \Big) \Big\{1+ o(1)\Big\}.$$
\noindent
Combining the above results, we  immediately obtain the MSE for all the elements in $\hat \Sigma_{\tiny\mbox{SDB}}$ as $\mbox{MSE}(\hat{\mbox{SE}}^2_{\tiny\mbox{SDB}}) =(  c_1/N^2R+ c_3/N^2n^2 +c_2/N^3   +3c_4/N^2nR) \{1+ o(1)\}$,  where  $c_4= \sum_{j_1\neq j_2} (\sigma_{j_1,j_2,2}-\sigma_{j_1,j_1}\sigma_{j_2,j_2})$. When  $R$ is of a larger order of $n$,
we have
\beqr
\mbox{MSE}(\hat{\mbox{SE}}^2_{\tiny\mbox{SDB}})=\Big(  \frac{c_1}{N^2R}+ \frac{c_3}{N^2n^2} +\frac{c_2}{N^3}   \Big) \Big\{1+ o(1)\Big\}.
\label{MSE_SDB}
\eeqr

We note that the MSEs of these bootstrap estimators in \eqref{MSE_BLB}, \eqref{MSE_SB} and \eqref{MSE_SDB}
based on multivariate random variables are similar to their univariate counterparts but with different constants in the numerators.
\section{Optimal Subsampling Bootstraps and General Parameters}\label{sec: optimal}
\subsection{Optimal subsampling bootstraps}\label{sec:osb}
All the subsampling bootstrap methods discussed so far require the specification of the hyperparameters including the size of the subsamples $n$, the number of subsamples $R$, and for the BLB, the number of the resamples $B$. For these methods to produce satisfactory statistical performance, intuitively these hyperparameters should be set as large as possible. However, setting them unnecessarily large will incur additional computational cost that we want to avoid in the first place when resorting to subsampling bootstrap instead of the bootstrap. Thus, there is a clear trade-off between statistical accuracy of the estimator and computational budget. Our approach to achieve the optimal trade-off is to optimize the statistical accuracy subject to a fixed pre-specified computational constraint.

Having analyzed the MSE of the three subsampling bootstrap methods BLB, SB, and SDB in the previous section, we are ready to derive the hyperparameter values that give optimal statistical accuracy in terms of MSE with a fixed computational cost.
Since these bootstrap methods are solely developed for massive data, it makes sense to compare their performance when the full sample resides on a hard drive. That is, to conduct any bootstrap scheme, the data will have to be repeatedly read from the hard disk, while the computation happens in memory. 
Sampling $n$ data points from the disk costs $O(n)$ in terms of time complexity. Denote the computational time for obtaining an estimator on a sample with size $n$ as $t(n)$.  For an estimator taking weighted data representation, we assume that the estimation time is $t(n)$ where $n$ is the number of distinct points in the sample. This is the scenario where BLB and SDB can gain computationally over TB. We summarize the sampling  and  computational cost in Table \ref{tab:cost}, with more details provided when a specific method is discussed. Later in this paper, we focus on those estimators where $t(n)\asymp n $, that is, those estimators that can be obtained in a computational time linear in the sample size. This setup comprises a large class of estimators, including   the sample mean and the least-squares estimator of the coefficients in a linear model.  For example,  to  compute the mean discussed so far, the total time required for  BLB is $\alpha_1 \cdot nR + \alpha_2 \cdot nRB$, where $\alpha_1$ and $\alpha_2$ are two parameters specific to the computer system. See below for more details.

\begin{table}[htp!]
\caption{Sampling time and estimation time of various subsampling bootstrap methods. $N$: the total sample size; $n$: the sample size of subsamples; $R$: the number of subsamples; $B$: the number of resamples.}
\begin{center}
\begin{tabular}{|c|c|c|}\hline
Name & Sampling Time & Estimation Time\\\hline
Bootstrap & $NB$ & $(B+1)\times t(N)$\\
BLB & $nR$ & $R(B+1) \times t(n)$\\
SB & $nR$ &$R \times  t(n)$\\
SDB & $nR$ & $ 2R \times t(n)$\\\hline
\end{tabular}
\end{center}
\label{tab:cost}
\end{table}%

We start by looking at SB first.  Since  SB needs to  sample $n$ data points
 from the disk $R$ times, its sampling cost is $O(nR)$. To compute  $\hat{\mbox{SE}}^2_{\tiny\mbox{SB}}$, the leading term is $\bar X^{(b)}$, i.e., the sample average of each bootstrap subsample, which is evaluated $R$ times, resulting in a computational cost of $O(nR)$. Therefore, the total computational cost for the SB is of the order $O(nR)$. Motivated by \eqref{MSE_d} and \eqref{MSE_SB}, to identify the optimal subsample size and optimal number of subsamples in terms of minimizing the MSE of $\hat{\mbox{SE}}^2_{\tiny\mbox{SB}}$, we need to find the $(n, R)$ pair such that
 \[ (n^*, R^*) =\arg\min_{n, R} ~\mbox{MSE}(\hat{\mbox{SE}}^2_{\tiny\mbox{SB}}) = \arg\min_{n, R}~\left\{\frac{c_1'}{R}+\frac{c_2'}{n^2}\right\}, \quad s.t. \quad \alpha_{\mbox{\tiny SB}}  \cdot n  R \le C_{\max},
 \]
 where $\alpha_{\mbox{\tiny SB}}$ is a constant quantifying the time constant before $nR$ for  the  SB implementation,
 and $C_{\max}$ is the computational budget.   Here, $c_1'=2$ and $c_2'=1$ when estimating the mean parameter for univariate random variables. For multivariate random variables,  we have
  $c_1'=c_1$ and $c_2'=c_3$ where $c_1$ and $c_3$ are defined in \eqref{c1} and \eqref{c3} respectively. Note that for the multivariate case, $c_1$ and $c_3$ can be consistently estimated and the computational cost of obtaining them is negligible in comparison to the bootstrap procedure.
 In practice, if we have the computational  budget to subsample $R$ times, each subsample with size $n$,  that is, the computational budget is $nR$, then solving the above optimization problem gives SB optimal hyperparameters:
\begin{equation}
 n^*=\lfloor \{(2c'_2/c'_1)\cdot(C_{\max}/\alpha_{\mbox{\tiny SB}})\}^{1/3} \rfloor, \quad
  R^*= \lfloor (c'_2/2c'_1)^{1/3} \cdot  (C_{\max}/\alpha_{\mbox{\tiny SB}})^{2/3} \rfloor, \label{eq:sb}
\end{equation}
where  $ \lfloor s\rfloor$  stands for the largest integer  that is not larger than $s$.

For the SDB procedure, the MSEs for the univariate and multivariate case as in \eqref{MSE_e} and \eqref{MSE_SDB} are similar to those of the SB in that the optimal $(n, R)$ pair is given by SDB optimal hyperparameters:
\begin{equation}
 n^*=\lfloor \{(2c'_2/c'_1)\cdot(C_{\max}/\alpha_{\mbox{\tiny SDB}})\}^{1/3} \rfloor, \quad R^*= \lfloor (c'_2/2c'_1)^{1/3} \cdot  (C_{\max}/\alpha_{\mbox{\tiny SDB}})^{2/3} \rfloor.  \label{eq:sdb}
\end{equation}
In practice,  $\alpha_{\mbox{\tiny SB}}$ and $\alpha_{\mbox{\tiny SDB}}$ can be  dependent on the subsample size $n$ because different computer architectures implement their algorithms differently. However, their dependence on $R$ is linear because resampling is just a repetition of the same operation.  Our strategy is to tune  $\alpha_{\mbox{\tiny SB}}$ or $\alpha_{\mbox{\tiny SDB}}$ progressively several times. The time needed to tune them can be  costly  but we minimize its impact by pilot-running experiments on small $n$ and small $R$.  The details can be found in   the simulations in Section \ref{simu:lr} --\ref{simu:logit} and also  Appendix B.1.

We are now ready to discuss how to choose optimal hyperparameters for BLB which involves three parameters $n$, $R$ and $B$. The sampling cost of BLB is similar to that of SB which is $O(nR)$. The computational cost can be seen as $O(nRB)$ as the leading computation is to calculate  $\overline X^{(r,b)}$, each based on a resample of size $n$, for a total of $RB$ times. Motivated by the expression for the MSE in \eqref{MSE_c} and \eqref{MSE_BLB}, to identify the optimal hyperparameter triple, we minimize the MSE of $\hat{\mbox{SE}}^2_{\tiny\mbox{BLB}}$ subject to the constraint that the computational cost is capped at $C_{\max}$. More specifically, we solve $(n^*,  R^*, B^*)$ equals to
\begin{equation}
 \arg\min_{n, R, B} ~\mbox{MSE}(\hat{\mbox{SE}}^2_{\tiny\mbox{BLB}}) = \arg\min_{n, R, B}~\left\{\frac{\wt c_1}{RB}+\frac{ \wt c_2}{nR}+ \frac{\wt c_3}{n^2} \right\}, \\
s.t. ~~ \alpha_1  \cdot n RB+ \alpha_2 \cdot nR \le C_{\max}, \label{blb:opt}
\end{equation}
where  $\wt c_1=2 \sigma^4(\sigma_4-\sigma^4)^{-1}$, $\wt c_2=1$, and  $\wt c_3=\sigma^4(\sigma_4-\sigma^4)^{-1}$  for the univariate mean estimating case and $\wt c_1=c_1$, $\wt c_2=c_2$, and
 $\wt c_3=c_3$ for the multivariate case with $c_1$, $c_2$, and $c_3$ defined in \eqref{c1}, \eqref{c2}, and \eqref{c3}, respectively.
 In the above optimization problem, $\alpha_1$ and $\alpha_2$ are two constants quantifying system specific constants for the sampling cost and the computational cost respectively. To figure out their values for a specific problem, we can take different combinations of $(n, R, B)$ and record the running time of  BLB under these combinations. Fitting a linear model with the running time as the response variable and  $nRB$ and $nR$ as the two covariates, we can obtain the estimates of $\alpha_1$ and $\alpha_2$. Our simulation study below suggests that  with very few  combinations (e.g., 8), we can obtain fairly accurate estimates of $\alpha_1$ and $\alpha_2$ with the associated  R$^2$ as large as 98\%. In estimating $\alpha_1$ and $\alpha_2$ using pilot runs, we can keep $R$, the number of resamples, small because any dependence of the computational and sampling time on $R$ will be proportional to it. Therefore, the extra time needed to estimate $\alpha_1$ and $\alpha_2$ is negligible in comparison to the actual implementation of any bootstrap  subsampling method discussed so far because for any bootstrap, $R$ is often  very large.

 Solving the optimization problem above  for BLB amounts to finding  the minimum value of  $ \wt c_1(RB)^{-1}+ \wt c_2(nR)^{-1}+\wt c_3(n)^{-2} $ under the constraint  $\alpha_1 n RB+ \alpha_2 nR=C_{\max}$.
 To this end, let $(n, R, B)$ be an arbitrary specification such that  $\alpha_1 nRB+ \alpha_2 nR  = C_{\max}$.  By
  Cauchy's inequality, we have
\beqr
&& \Big( \frac{\wt c_1} {RB}+\frac{ \wt c_2}{nR}+ \frac{\wt c_3}{n^2} \Big) \Big(\alpha_1  n RB+ \alpha_2 nR\Big) \n\\
 = &&\Big(\frac{ \wt c_1} {RB}+\frac{\wt c_2}{nR} \Big) \Big(\alpha_1  n RB+ \alpha_2 nR \Big)+\frac{\wt c_3}{n^2} \Big(\alpha_1  n RB+ \alpha_2 nR \Big)
 \n \\
 \geq  && \Big(\sqrt {\wt c_1 \alpha_1 }  \sqrt {n}  + \sqrt{\wt c_2\alpha_2} \Big)^2 +  C_{\max} \frac{ \wt c_3 } {n^2}
:= f(n).
 \label{fn}
\eeqr
The  first  inequality  in \eqref{fn} becomes an equality  if
 $\wt  c_1\alpha_2 n/B = \wt c_2\alpha_1 B$. When $n$ is fixed,
this  leads to the BLB optimal hyperparameters as
\beqr
   R^*= \lfloor C_{\max}/(\alpha_1 n B^* +\alpha_2 n)\rfloor,  B^*=\lfloor  (\wt c_1 \alpha_2 /\wt c_2\alpha_1)^{1/2} \sqrt{n}\rfloor.
   \label{eq:blb}
\eeqr
For $n$, our simulation results suggest that  $n=\lfloor N^{0.7}\rfloor$   is an optimal choice in  most cases consistent with the findings in \cite{kleiner2014scalable} and \cite{sengupta2016subsampled}. Accordingly, we can directly use $n=\lfloor N^{0.7}\rfloor$  in practice.

\subsection{General parameters and statistics }\label{sec:general}
We have so far focused on estimating the mean. In this subsection, we study  more general parameters.
 Without loss of generality, we first assume  that the random variables $X_i\in \mR^1, i=1, ..., N,$ have mean
 $\mu$ and variance $\sigma^2$.  Our interest is to estimate $\theta=g(\mu)$,
 where $g(\cdot)$ is a known, possibly complicated, but  sufficiently smooth function with twice differentiable. Extension to random vectors is straightforward.
One simple estimator of $\theta$ can be constructed as $\hat \theta =g(\overline X)$ since $\overline X$ estimates $\mu$. Asymptotically,
 we have $\hat \theta-\theta= \dot g(\mu)(\overline X-\mu)\{1+o_p(1)\}$,
 where  $\dot g(\cdot)$ is the first order derivative of
 $g(\cdot)$.
  We then have $\sqrt N(\hat \theta-\theta)\rightarrow_d N(0, \dot g^2(\mu)\sigma^2)$.
  This means that the asymptotic SE$^2$  of  $\hat \theta$ is given by
 $ \mbox{SE}^{*2}=\dot g^2(\mu) \sigma^2/N. $
 Accordingly, a natural estimator for $ \mbox{SE}^{*2}$  is given by
 $\wh {\mbox{SE}}^{*2}=\dot g^2(\overline X) \hat \sigma^2/N, $
 which can be easily  computed if $\dot g^2(\mu)$ is analytically simple.

 Next, we discuss how to use the three subsampling bootstraps discussed so far to estimate SE$^{*2}$.
 We start with the traditional bootstrap method for which a natural estimator of $ \mbox{SE}^{*2}$  is given by
 \beqr
 {\wh {\mbox{SE}}^{*2}}_{\tiny \mbox{TB}} &=& B^{-1} \sum_{b=1}^{B}\Big\{g(\overline X^{(b)})- g(\overline X)\Big\}^2. \label{SE_B*}
 \eeqr
 It is seen that
 \beqr
  {\wh {\mbox{SE}}^{*2}}_{\tiny \mbox{TB}}
 = B^{-1} \sum_{b=1}^{B} \Bigg\{ \dot g(\overline X)\Big(\overline X^{(b)}-\overline X\Big) +2^{-1} \ddot g^2(\overline X^*)\Big(\overline X^{(b)}-\overline X\Big)^2 \Bigg\}^2  \n
 \eeqr
\[  =B^{-1} \sum_{b=1}^{B}\Big\{\dot g^2(\mu) \big(\overline X^{(b)}-\overline X\big)^2+\dot g(\mu)
\ddot g^2(\mu)\big(\overline X^{(b)}-\overline X\big)^3 +4^{-1}\ddot g^4(\mu)\big(\overline X^{(b)}-\overline X\big)^4\Big\}\Big\{1+o_p(1)\Big\},\]
where   $\ddot g(\cdot)$ is the second order derivative of
 $g(\cdot)$ and $\overline X^*$ is between $\overline X$ and $\overline X^{(b)}$.
Denote
${\wh {\mbox{SE}}^{*2}}_{\tiny \mbox{TB}}
 =  (\dot g^2(\mu) A^*_1+\dot g(\mu)
\ddot g^2(\mu) A^*_2+ 4^{-1}\ddot g^4(\mu)A^*_3)\{1+o_p(1)\}.$
According to the results in Section 1.10 of the Supplementary Material, we have MSE$(A^*_2)=O((N^3B)^{-1})$.
If MSE$(A^*_2)$ is ignorable as compared with  MSE$(A^*_1)$, it  can be verified that    MSE$({\wh {\mbox{SE}}^{*2}}_{\tiny \mbox{TB}})$  should be determined by  MSE$(A^*_1)$ and the reminder terms can be ignored accordingly.
By the result of Theorem 2, we have MSE$(A^*_1) =\mbox{MSE}({\wh {\mbox{SE}}^{2}}_{\tiny \mbox{TB}}) =O(N^{-3}).$
Hence,  we  have
\[ {\wh {\mbox{SE}}^{*2}}_{\tiny \mbox{TB}}
 =  \dot g^2(\mu) A^*_1\Big\{1+o_p(1)\Big\} = \dot g^2(\mu)  \Big\{B^{-1} \sum_{b=1}^{B}  \big(\overline X^{(b)}-\overline X\big)^2\Big\}\Big\{1+o_p(1)\Big\}
 =\dot g^2(\mu) {\wh {\mbox{SE}}^{2}}_{\tiny \mbox{TB}}\Big\{1+o_p(1)\Big\}.\]
 This verifies the fact that the asymptotic behavior of ${\wh {\mbox{SE}}^{*2}}_{\tiny \mbox{TB}} $ is  determined by that of ${\wh {\mbox{SE}}^{2}}_{\tiny \mbox{TB}} $  defined in Section 2.1 and higher order terms  can be ignored. Hence, $c_1'$, $c_2'$ and $c_3'$ are  the same constants  defined  for the sample mean and can be consistently estimated.
Similar analysis can be performed for the BLB, SB, and SDB methods, demonstrating that higher order terms can be ignored.
The associated estimators  for     $ \mbox{SE}^{*2}$   are given by
 \beqr
{\wh {\mbox{SE}}^{*2}}_{\tiny \mbox{BLB}}&=&  (RB)^{-1} \sum_{r=1}^R   \sum_{b=1}^{ B} \Big\{g(\overline X^{(r,b)})- g(\overline X^{(r)})\Big\}^2= \dot g^2(\mu)\wh {\mbox{SE}}_{\tiny \mbox{BLB}}^2 \Big\{1+o_p(1)\Big\},\label{SE_C*}
\\
 {\wh {\mbox{SE}}^{*2}}_{\tiny \mbox{SB}}&=& \Big(\frac{n}{N}\Big){R}^{-1} \sum_{r=1}^{R} \Big\{g(\overline X^{(b)})-g(\overline X)\Big\}^2
=  \dot g^2(\mu)\wh {\mbox{SE}}_{\tiny \mbox{SB}}^2 \Big\{1+o_p(1)\Big\}, \label{SE_D*}
\\
{\wh {\mbox{SE}}^{*2}}_{\tiny \mbox{SDB}}&=& R^{-1} \sum_{r=1}^R    \Big\{ g(\overline X^{(r,1)})-g(\overline X^{(r)})\Big\}^2
=  \dot g^2(\mu)\wh {\mbox{SE}}_{\tiny \mbox{SDB}}^2 \Big\{1+o_p(1)\Big\},  \label{SE_E*}
\eeqr
respectively.

  When $X_i\in \mR^p$ is a  random vector   and $\theta$ is one-dimensional, we have $\hat \theta-\theta=g(\overline X)-g(\mu)= \dot g(\mu)^\top(\overline X-\mu)\{1+o_p(1)\}$, leading to
$\sqrt N(\hat \theta-\theta)\rightarrow_d N(0,  \dot g(\mu)^\top\Sigma \dot g(\mu))$.
 This means that the asymptotic SE$^2$  of  $\hat \theta$ is given by
 $ \mbox{SE}^{*2}= \dot g(\mu)^\top \Sigma \dot g(\mu)/N $.
 Similar to the above calculation,  we have
 ${\wh {\mbox{SE}}^{*2}}= N^{-1}\dot g(\mu)^\top\hat \Sigma\dot g(\mu)\{1+o_p(1)\}$
 and  $\wh {\mbox{SE}}^{*2} -\mbox{SE}^{*2}= N^{-1}\dot g(\mu)^\top(\hat \Sigma-\Sigma) \dot g(\mu)\{1+o_p(1)\}.$
 This suggests that  the asymptotic behavior of ${\wh {\mbox{SE}}^{*2}}$ is   determined by the estimate  of $\Sigma$.

 When  $\theta\in \mR^d$ for $d>1$, we use $\Sigma^*$ to denote the covariance matrix of $\hat \theta=g(\overline X)$. By a careful calculation, we can find that   $\Sigma^*=\dot g(\mu)^\top \Sigma \dot g(\mu)/N$  still holds.
We then discuss   how to use the three subsampling bootstraps to estimate  $ \Sigma^*$.
 We first study  the SB method for which a natural estimator  is given by
 \beqr
{ \wh {\Sigma}^*}_{\tiny \mbox{SB}} &=& n(NR)^{-1} \sum_{r=1}^{R}\Big\{g(\overline X^{(r)})- g(\overline X)\Big\}\Big\{g(\overline X^{(r)})- g(\overline X)\Big\}^\top\n\\
 &=& \Bigg\{n(NR)^{-1}\sum_{r=1}^{R} \dot g(\mu)^\top\Big(\overline X^{(r)}-\overline X\Big)\Big(\overline X^{(r)}-\overline X\Big)^\top g(\mu) \Bigg\} \Big\{1+o_p(1)\Big\}\n\\
&=& \dot g(\mu)^\top {\wh \Sigma}_{\tiny \mbox{SB}}\dot g(\mu)\Big\{1+o_p(1)\Big\}. \n
\eeqr
This suggests that the asymptotic behavior of  $ \wh\Sigma^*_{\tiny \mbox{SB}}$ is  largely determined by that of
 $\wh {\Sigma}_{\tiny \mbox{SB}}$.
Similar analysis can be  conducted for the BLB and SDB  methods. The associated estimators  for     $ \Sigma^*$   are given by
 \[
{\wh {\Sigma}^*}_{\tiny \mbox{BLB}}=  (RB)^{-1} \sum_{r=1}^R   \sum_{b=1}^{ B} \Big\{g(\overline X^{(r,b)})- g(\overline X^{(r)})\Big\}\Big\{g(\overline X^{(r,b)})- g(\overline X^{(r)})\Big\}^\top =\dot g(\mu)^\top \wh {\Sigma}_{\tiny \mbox{BLB}}\dot g(\mu) \Big\{1+o_p(1)\Big\},\]
\[{\wh {\Sigma}^*}_{\tiny \mbox{SDB}}= R^{-1} \sum_{r=1}^R    \Big\{ g(\overline X^{(r,1)})-g(\overline X^{(r)})\Big\}\Big\{ g(\overline X^{(r,1)})-g(\overline X^{(r)})\Big\}^\top
=  \dot g(\mu)^\top \wh {\Sigma}_{\tiny \mbox{SDB}} \dot g(\mu) \Big\{1+o_p(1)\Big\},  \]
respectively. Again, the asymptotic behaviors of ${\wh {\Sigma}^*}_{\tiny \mbox{BLB}}$, ${\wh {\Sigma}^*}_{\tiny \mbox{SB}}$, and ${\wh {\Sigma}^*}_{\tiny \mbox{SDB}}$  are   largely determined by those of  ${\wh {\Sigma}}_{\tiny \mbox{BLB}}$, ${\wh {\Sigma}}_{\tiny \mbox{SB}}$, and ${\wh {\Sigma}}_{\tiny \mbox{SDB}}$.  Accordingly, the above mentioned optimization approaches in Section \ref{sec:osb} are still applicable  for the  general  parameters and statistics, as long as the target parameter can be expressed  as $\theta= g(\mu)$ where $g(\cdot)$ is a sufficiently smooth function.

Note that the computational advantages of SDB and BLB relative to the traditional bootstrap lies in the fact that the estimator of interest can take weighted data representation as its argument. A large class of commonly used estimators, including M-estimators,  can be represented by weighted data.
We now discuss a few examples that can be placed in the general parameter framework.

{\bf Example 1 (Estimators based on moments)}.  Denote $\mu_{j}=E(X-\mu)^{j}$ as the $j$th moment  of a univariate random variable $X$ and its moment estimator as $\hat \mu_{j}=N^{-1}\sum_{i=1}^{N}(X_i-\bar X)^{j}$. Note that $\hat \mu_j$ is a consistent estimator for $\mu_j$ which can take weighted data representation. Thus, the optimal hyperparameter selection method presented in Section \ref{sec:osb}  is  applicable for minimizing the MSE of $\hat \theta=g(\hat \mu)$ when one uses the subsampling bootstrap for inference.

As another example,  we will later apply our method to the problem of estimating the correlation coefficient when missing data is present, a problem considered in  \cite{shao2002sample}. Denote the dataset as  $X=(X_1,\cdots,X_N)\in\mR^N$ and $Y=(Y_1,\cdots,Y_N)\in\mR^N$ and the missing indicator as   $w_i$, where $w_i=1$  if the $i$ th observation  $(X_i, Y_i)$ is missed and $w_i=0$ if not.   \cite{shao2002sample} is interested in estimating the correlation coefficient
\[ \rho=\frac{\mu_{\scriptscriptstyle{XY}}-\mu_{\scriptscriptstyle X}\mu_{\scriptscriptstyle Y}}{\sqrt{\mu_{\scriptscriptstyle{X,2}}-\mu_{ \scriptscriptstyle X}^2}
\sqrt{\mu_{\scriptscriptstyle{Y,2}}-\mu_{ \scriptscriptstyle Y}^2}},
\]
where $\mu_{\scriptscriptstyle X}= E(w_i X_i)$, $\mu_{\scriptscriptstyle  {X,2}}= E(w_iX_i^2)$, $\mu_{\scriptscriptstyle Y}= E(w_i Y_i)$,  $\mu_{\scriptscriptstyle {Y,2}}= E(w_iY_i^2)$,  and
 $\mu_{\scriptscriptstyle{XY}}= E(w_iX_iY_i)$.  A simple sample estimator is given by
\beqr
\hat \rho= \frac{\sum_{i=1}^{N}w_iX_iY_i-\hat X\hat Y/\hat N}{\sqrt{\sum_{i=1}^{N}(w_i X_i^2-\hat X^2)}\sqrt{ \sum_{i=1}^{N}(w_i Y_i^2-\hat Y^2)}},\label{sample:corr}
\eeqr
which  satisfy $\hat{\rho}=\rho+o(1)$. Write $\mu=( \mu_{\scriptscriptstyle X}, \mu_{\scriptscriptstyle  {X,2}}, \mu_{\scriptscriptstyle Y}, \mu_{\scriptscriptstyle  {Y,2}}, \mu_{\scriptscriptstyle  {XY}})^\top$. Then we can write $\rho= g( \mu)$ and $\hat \rho= g( \hat\mu)$. The   optimization technique for multivariate random variables  in Section \ref{sec:osb} can be directly applied to optimize the estimation accuracy  of  SE$^2$ for  $\hat \rho$.

{\bf Example 2 (Linear regression)}. Consider the problem of estimating the regression coefficient in linear regression $Y_i=X_i^\top\beta +\ve_i$, $i=1,\cdots, N$ where $X_i\in\mR^{d}$ and  $\beta\in\mR^d$ is  the regression coefficient. We use $d=2$ for illustration purposes. Note the ordinary least-squares estimator of  $\beta$ is given by
 $\hat \beta= (X^\top X)^{-1} X^\top Y$.  Write   $\mu=(\mu_{1}, \mu_{2}, \mu_{3}, \mu_{4}, \mu_{5})^\top$ where  $\mu_{1}= E(X^2_{i,1})$, $\mu_{2}= E(X^2_{i,2})$, $\mu_{3}= E(X_{i,1}X_{i,2})$, $\mu_{4}= E(X_{i,1} Y_i)$ and $\mu_{5}= E(X_{i,2} Y_i)$. It is easy to see $\hat \beta =g(\hat \mu)$ for some function $g$.
Since each element of  $\hat \mu$ is  a mean estimator, we can apply the results in Section \ref{sec:osb} to choose the optimal choice for various hyperparameters.

{\bf Example 3 (Logistic regression)}. Consider the problem of estimating the regression coefficient in logistic regression $P(Y_i=1|X_i)=p_i(\beta)=\exp(X_i^\top \beta)/\{1+\exp(X_i^\top \beta)\}$, $i=1,\cdots, N$ where $X_i\in\mR^{d}$ is the covariate  and  $\beta\in\mR^d$ is  the regression coefficient. We still  use $d=2$ for illustration. The one-step  estimator of  $\beta$ is given by
 $\hat \beta= \wt \beta+ \{\sum_{i=1}^N \omega_i(\wt \beta) X_iX_i^\top\}^{-1}\sum_{i=1}^N\{Y_i-p_i(\wt\beta)\}X_i$
 with $\omega_i(\beta)=p_i(\beta)\{1-p_i(\beta)\}$, where $\wt \beta$ is a  pilot, consistent estimator of $\beta$ using a subsample with sample size $n$.  Write   $\mu=(\mu_{1}, \mu_{2}, \mu_{3}, \mu_{4}, \mu_{5})^\top$ where  $\mu_{1}= E(\omega_i(\beta)X^2_{i,1})$, $\mu_{2}= E(\omega_i(\beta)X^2_{i,2})$, $\mu_{3}= E(\omega_i(\beta)X_{i,1}X_{i,2})$, $\mu_{4}= E[X_{i,1}\{Y_i-p_i(\beta)\}]$ and $\mu_{5}= E[X_{i,2}\{Y_i-p_i(\beta)\}]$. Then  $\hat \beta =g(\hat \mu)$ for some function $g$ when $\beta$ is replaced by $\wt\beta$.

 {\bf Example 4 (Two stage least squares)}.  Consider the problem of estimating the regression coefficient in linear regression $Y_i=X_i\beta +u_i$, $i=1,\cdots, N$, where the random error $u_i$ is correlated with $X_i\in \mR^1$ such that $\cov(X_i,u_i)\neq 0$.  To estimate $\beta$, an idea is to identify an instrumental variable, denoted as $Z_i\in\mR^1$, and employ the so-called two stage least squares estimator given by
 $\hat \beta_{IV}=  (\sum_{i=1}^N Z_iY_i)/(\sum_{i=1}^N Z_iX_i)$.
   Write   $\mu=(\mu_{1}, \mu_{2})^\top$ where  $\mu_{1}= E(Z_i Y_i)$ and  $\mu_{2}= E(Z_iX_i)$.  Clearly,  we have  $\hat \beta =g(\hat \mu)$ for some function $g$.

\section{ Numerical Study}
We conduct extensive simulation study to assess the performance of the hyperparameter specification method. We will do this by first verifying the MSE results in Section \ref{simu:mse} for estimating the mean.  In Section \ref{simu:lr}, we  apply our hyperparameter selection method to linear regression and further
 extend it to logistic regression  in Section \ref{simu:logit}.   Due to page limit, we leave the estimation of the correlation coefficient defined for missing data, hypotheses test and comparison with plain subsampling to the supplementary Material.
  All programs
are written in python 3.7 and run on a  cloud computing platform called Matpool (https://www.matpool.com). The simulation is run on a computer
 equipped with NVIDIA Tesla P100  data  center GPU, 16 GB of graphics  memory, 64 GB ram and 500 GB SSD capacity.

\subsection{Verify the formula for the  mean squared errors}  \label{simu:mse}
This section aims to verify the MSE($\hat{\mbox{SE}}^2$) formula derived in Section 2 as expressed in \eqref{MSE_a},  \eqref{MSE_b}, \eqref{MSE_c}, \eqref{MSE_d}, and  \eqref{MSE_e}, when the mean is estimated. Towards this, we repeat each experiment $M$ times  and denote $ \hat {\mbox{SE}}^{2(m)}$ as the  estimate of SE$^2$ on the $m$th simulation replicate. The true mean squared error (MSE)  can  then be estimated as
 \beqr
\hat{\mbox{MSE}}= M^{-1}\sum_{m=1}^M \Big(\hat {\mbox{SE}}^{2(m)}-{\mbox{SE}}^{*2}\Big)^2
\label{MSE_cp}
 \eeqr
 where $\mbox{SE}^{*2} = M^{-1}\sum_{m=1}^M(\overline X^{(m)}-\overline X)^2$ and
 $\overline X^{(m)}$ is the estimator of $\overline X$  based on the $m$th simulation replicate. With some abuse of notation, denote  $\mbox{MSE}^{*}$ as the estimator of MSE($\hat{\mbox{SE}}^2$) in \eqref{MSE_a},  \eqref{MSE_b}, \eqref{MSE_c}, \eqref{MSE_d}, and  \eqref{MSE_e},  where
 $\sigma$ and $\sigma_4$ are estimated as  $\hat \sigma^2= \sum_{i=1}^{N} (X_i-\overline X)^2/N$ and $\hat \sigma_4= \sum_{i=1}^{N}(X_i-\overline X)^4/N$, respectively.   We then compare the difference between  $\hat {\mbox{MSE}}$  and  $ \mbox{MSE}^{*}$ by evaluating  the ratio $\mbox{MSE}^{*}/\hat{\mbox{MSE}}$.

 We generate $X_i\in\mR^1$ for $i=1,\cdots,N$ either from a standard normal distribution (Normal) or a centered standard exponential distribution (Exponential).
 We take $N=10^5, n\in \{\lfloor N^{0.4}\rfloor,  \lfloor N^{0.5}\rfloor, \lfloor N^{0.6}\rfloor\}$, $B\in \{25,50\}$, and $R\in \{25,50\}$. For each setup, a total of $M=1000$ datasets are generated. The simulation results are summarized in Table \ref{tab:t2}.  From this table, we observe that all of the ratios are close to one, suggesting that our analytical formula are fairly accurate.  Similarly, we can verify the formula for  MSE($\hat{\mbox{SE}}^2$) derived in \eqref{MSE_BLB},  \eqref{MSE_SB},  and \eqref{MSE_SDB} with $X_i\in\mR^p$.  We find that the results are similar to those presented in Table \ref{tab:t2} and omit them to save space.

 \begin{table}[!hbt]
\bc\emph{}
\caption{\label{tab:t2} Simulation results for checking the analytical formula of MSE for various bootstrap methods. }
{\small
\begin{tabular}{ccc|Hcccc|Hcccc}
\hline
\hline
 \multicolumn{3}{c}{Parameter} & \multicolumn{5}{c}{Normal}  & \multicolumn{5}{c}{Exponential}\\
   $n$ & $B$&   $R$&AF   &  TB&BLB   &   SB&SDB&AF     &  TB&BLB   &   SB&SDB\\
  \hline
   $ \lfloor N^{0.4} \rfloor$&    25 &    25 &     1.007   &     0.982 &     0.979 &     1.004 &     1.002 &     1.055 &     1.003 &     0.969 &     0.944 &     1.116  \\
                            &    25 &    50 &      1.007 &     0.982 &     0.914 &     1.020 &     1.010 &     1.055 &     1.003 &     1.001 &     0.976 &     1.061  \\
                             &   50  &    25 &   1.007 &     0.927 &     0.975 &     1.004 &     1.002 &     1.055 &     1.006 &     0.929 &     0.944 &     1.116   \\
                             &    50 &    50 &     1.007   &     0.927   &     0.987   &      1.020  &       1.010 &       1.055 &       1.006 &       1.023 &       0.976 &     1.061   \\
                             \hline
     $ \lfloor N^{0.5} \rfloor$ &   25 &   25 &       1.007 &     0.982 &     1.042 &     1.075 &     1.000 &     1.055 &     1.003 &     0.991 &     0.964 &     1.069  \\
                                  &    25 &    50 &     1.007 &     0.982 &     0.982 &     1.075 &     1.014 &     1.055 &     1.003 &     1.028 &      1.013 &     1.032   \\
                                  &    50 &    25 &     1.007 &     0.927 &     0.937 &     1.014 &     1.000 &     1.055 &     1.006 &     0.962 &     0.964 &     1.069   \\
                                   &    50 &    50 &     1.007 &     0.927 &     0.950 &     1.014 &     1.014 &     1.055 &     1.006 &     0.946 &     1.013 &     1.032   \\\hline

  $ \lfloor N^{0.6} \rfloor$ &    25 &    25 &     1.007 &     0.982 &     0.946 &     0.877 &     0.932 &     1.055 &     1.003 &     1.044 &     0.967 &     0.951  \\
                               &    25 &    50 &     1.007 &     0.982 &     1.003 &     0.877 &     1.030 &     1.055 &     1.003 &     0.975 &     1.121 &     0.977   \\
                                  &    50 &    25 &     1.007 &     0.927 &     1.059 &     0.991 &     0.932 &     1.055 &     1.006 &     0.965 &     0.967 &     0.951   \\
                                   &    50 &    50 &     1.007 &     0.927 &     1.082 &     0.991 &     1.030 &     1.055 &     1.006 &     1.049 &     1.121 &     0.977   \\
                         \hline
   \end{tabular}}
\ec
\end{table}

 \subsection{Hyperparameter selection for  linear regression} \label{simu:lr}
As an application, we
consider  a two-dimensional linear  regression model  $Y_i=X_i^T\beta+\ve_i$ for $i=1,\cdots, N$, where $\beta_0=(0.1,0.1)^\top$ is the regression coefficient to be estimated, $X_i$ follows a bivariate normal distribution $N(0,\Sigma)$ with $\Sigma=(1, 0; 0, 1)$ and $\ve_i$ are the independent noise following the standard normal distribution. We denote the ordinary least-squares estimator as $\hat\beta$ whose theoretical covariance is
$ \Cov(\hat\beta)=\Sigma/N$.
   This covariance  can be consistently estimated by the BLB as
\[\hat {\mbox{SE}}^{*2}_{\tiny\mbox{BLB}}= \frac{1}{RB} \sum_{r=1}^R   \sum_{b=1}^{ B} (\hat \beta^{(r,b)}- \hat \beta^{(r)}) (\hat \beta^{(r,b)}- \hat \beta^{(r)})^\top,\]
 where  $\hat \beta^{(r,b)}$ is the ordinary least-squares  estimator obtained from the $b$th resample of the $r$th subsample and $\hat  \beta^{(r)}$ is the ordinary least-squares estimate based on the $r$th resample. Since we are estimating matrices,  we define  $\mbox{MSE}(\hat {\mbox{SE}}^{*2}_{\tiny\mbox{BLB}})=\|\hat {\mbox{SE}}^{*2}_{\tiny\mbox{BLB}}-\Cov(\hat\beta)\|_F$ as the mean square error of the BLB estimate of $\Cov(\hat\beta)$. We can define the MSE of the SB and SDB methods likewise. In the experiments below, we fix $n=\lfloor N^{0.7}\rfloor$ for the prespecified setting and repeat the simulation under each setting $M=50$ times. To compare the performance of BLB, SDB and SB, we fix the computational budget at the same value of $C_{\max}$.  We emphasize  that through all of our simulations, for optimal BLB, the actual time budget is  always $C_{\max}-C_0$, where  $C_0$ is the cost for estimating  coefficients $\alpha_1$ and $\alpha_2$.

\begin{table}[!h]
\bc\emph{}
\caption{\label{tab:t6} Simulation results  for linear regression. Here, $(R,B)$ is the original specified setting for BLB with fixed  $n=\lfloor N^{0.7}\rfloor$.  We denote the BLB method using the specified $(R, B)$ combination as BLB, that using the optimal $(R,B)$ combination via our approach as BLB$^*$.
Denote $\kappa_1=\mbox{ BLB}^*$/ BLB, $\kappa_2=\mbox{ SDB}^*$/SDB,$\kappa_3=\mbox{ SB}^*$/SB,
$\kappa_4=\mbox{ BLB}^*/\mbox{ SDB}^*$, and $\kappa_5=\mbox{ BLB}^*/\mbox{ SB}^*$.}
\vspace{0.1cm}
{\scriptsize
\begin{tabular}{c|ccccc|ccccc}
\hline
\hline
 \multicolumn{1}{c}{Original} &  \multicolumn{5}{c}{MSE Ratios}  &  \multicolumn{5}{c}{CPU Time Ratios}\\
  ($R$, $B$) & $\kappa_1$&
   $\kappa_2$ & $\kappa_3$&$\kappa_4$&  $\kappa_5$&$\kappa_1$&
   $\kappa_2$ & $\kappa_3$&$\kappa_4$&  $\kappa_5$\\
\hline
(14,4089)&0.705 &0.276 & 0.780 & 0.580 & 0.208 & 0.995 & 1.075 &1.063 &      0.949 &      0.946   \\
(15,4097)&0.763 &0.289 & 0.745 & 0.502 & 0.194 & 0.986 & 1.095 &1.074 &      0.931 &      0.929   \\
(18,4291)&0.682 &0.257 & 0.718 & 0.551 & 0.206 & 1.002 & 1.035 &1.013 &      0.969 &      0.975 \\
(24,3847)&0.694 &0.259 & 0.779 & 0.593 & 0.220 & 0.996 & 1.076 &1.063 &      0.935 &      0.931   \\
(27,4430)&0.781 &0.319 & 0.816 & 0.569 & 0.214 & 1.003 & 1.020 &1.000 &      0.977 &      0.981   \\
(29,4205)&0.774 &0.294 & 0.770 & 0.588 & 0.216 & 0.997 & 1.072 &1.054 &      0.938 &      0.940 \\
  \hline
(363,65) & 0.865 &  0.276 &  0.821 &  0.600 &  0.223 &  0.984 & 1.071 & 1.044 & 0.930 &      0.947   \\
(408,57) & 0.799 &  0.262 &  0.689 &  0.571 &  0.213 &  0.980 & 1.057 & 1.029 & 0.938 &      0.956   \\
(414,71) & 0.810 &  0.283 &  0.809 &  0.579 &  0.220 &  0.978 & 1.018 & 0.993 & 0.962 &      0.979   \\
(471,66) & 0.822 &  0.307 &  0.839 &  0.521 &  0.200 &  0.986 & 0.999 & 0.976 & 0.994 &      1.009   \\
(513,67) & 0.804 &  0.300 &  0.786 &  0.545 &  0.189 &  0.965 & 0.998 & 0.969 & 0.960 &      1.074   \\
(584,68) & 0.666 &  0.293 &  0.981 &  0.488 &  0.166 &  0.953 & 0.960 & 0.945 & 0.984 &      1.097 \\
\hline
  \end{tabular}}
\ec
\end{table}

For this example, we evaluate the relative performance of BLB, SB and SDB under the same computational budget.  This is not always done in the literature due to the difficulty of controlling different bootstrap methods under the same time cost.  Here, the original SDB and SB corresponds to the cases with $n=\lfloor N^{0.7}\rfloor$ and $R$ is further calculated based on the time consumption of BLB.
In order to do this, we will need to estimate time coefficients $\alpha_{\mbox{\tiny SB}}$, $\alpha_{\mbox{\tiny SDB}}$, and $\alpha_1$ and $\alpha_2$ in the time constraints in the optimization problems in Section \ref{sec: optimal}, using pilot runs.

In our pilot runs, we randomly generate  12   combinations of $(R, B)$    values from  $\lfloor  U(1,10)\rfloor \times \lfloor  U(1,80)\rfloor $ and record the  computational time for each combination. Afterwards, we fit a linear regression model  with the running time as the response variable and  the corresponding $(nRB, nR)$ as the covariates for BLB. This gives ordinary least-squares estimates of $\alpha_1$ and $\alpha_2$.  For SB and SDB,   we use the iterative strategy described in Appendix B to estimate $\alpha$.

We can now compare BLB, SB and SDB with some initial specification of their hyperparameters. Towards this,
we  randomly generate  6 different combinations of ($R, B$) values    from   $ \lfloor U(15,30)\rfloor \times \lfloor U(2500,5000)\rfloor $  and another  6 pairs of ($R, B$) from $ \lfloor U(300,800)\rfloor \times \lfloor U(50,80)\rfloor $,  as seen in Table \ref{tab:t6}.   Their optimal specifications are obtained via  \eqref{eq:sb}, \eqref{eq:sdb} and \eqref{eq:blb}  under the multivariate case with $c_1'=c_1$ and   $c_2'=c_2$  for BLB
and  $c_1'=c_1$ and   $c_2'=c_3$ for SDB and SB, where $c_1$, $c_2$, and $c_3$ are defined in  \eqref{c1}, \eqref{c2}, and \eqref{c3}, respectively.
From this table,
we can see clearly that the running times of the optimal BLB, SB and SDB are all similar to those of the original setting with prespecified hyperparameters. On the other hand, all three methods with optimally chosen hyperparameters outperform their original settings. Moreover, SB always performs the worst consistent with the results in \cite{kleiner2014scalable}  and \cite{sengupta2016subsampled}.  It is also interesting that the optimal BLB  outperforms optimal SDB under the same time cost and the average improve margin is 44.3\%.

\subsection{Hyperparameter selection for  logistic regression} \label{simu:logit}
We now
consider  a logistic  regression model  by following the notation in Example 3 in Section \ref{sec:general}. The parameter to be estimated is set as $\beta_0=(0.5,0.5)^\top$ and we generate $X_i$  as in the linear regression model in Section \ref{simu:lr}. The covariance of the one-step estimator $\hat\beta$ can be consistently estimated by the BLB as
\[\hat {\mbox{SE}}^{*2}_{\tiny\mbox{BLB}}= \frac{1}{R B} \sum_{r=1}^{R}   \sum_{b=1}^{ B} (\hat \beta^{(r,b)}- \hat \beta^{(r)}) (\hat \beta^{(r,b)}- \hat \beta^{(r)})^\top,\]
where $\hat \beta^{(r,b)}$ is the one-step  estimator obtained from the $b$th resample of the $r$th subsample and $\hat  \beta^{(r)}$ is the one-step  estimate based on the $r$th resample. The pilot estimator $\tilde{\beta}$ is estimated using $n=\lfloor N^{0.7}\rfloor$ samples only once for each resample. The accuracy of the covariance estimator is evaluated by computing $\mbox{MSE}(\hat {\mbox{SE}}^{*2}_{\tiny\mbox{BLB}})=\|\hat {\mbox{SE}}^{*2}_{\tiny\mbox{BLB}}-\hat{\Cov}(\hat\beta)\|_F$, where for logistic regression, $\Cov(\hat\beta)=E\{ \sum_{i=1}^N\omega_i(\hat\beta) X_i X_i^\top \}^{-1}$ and $\hat{\Cov}(\hat\beta)=M_0^{-1}\sum_{i=1}^{M_0} \{\sum_{i=1}^N  \omega_i(\hat\beta) X^{(m)}_i X^{(m)\top}_i\}^{-1}$,
where $M_0=2000$ and $ X^{(m)}_i$ represents that $ X_i$ is generated in the $m$ th replication.
We can calculate  the MSE of the SB and SDB methods likewise.


\begin{table}[!hbt]
\bc\emph{}
\caption{\label{tab:t7} Simulation results  for logistic regression. The explanation
of the column headings can be found in the caption of Table \ref{tab:t6}. }
\vspace{0.1cm}
{\small
\begin{tabular}{c|ccc|ccc}
\hline
\hline
 \multicolumn{1}{c}{Original} &  \multicolumn{3}{c}{MSE Ratios}  &  \multicolumn{3}{c}{CPU Time Ratios}\\
  ($R$, $B$)& $\mbox{BLB}^*$/BLB&
   $\mbox{SDB}^*$/SDB & $\mbox{BLB}^*$/$\mbox{SDB}^*$ &$\mbox{BLB}^*$/BLB& $\mbox{SDB}^*$/SDB&  $\mbox{BLB}^*$/$\mbox{SDB}^* $ \\
\hline
   (20,3862) &      0.868 &      0.255 &      0.601 &      0.997 &      1.058 &      0.988   \\
   (21,3872) &      0.845 &      0.311 &      0.457 &     1.003 &      1.019 &      1.040   \\
   (22,4114) &      0.765 &      0.286 &      0.491 &       0.991 &      1.013 &      0.912   \\
    (26,3559) &      0.752 &      0.265 &      0.506 &   0.999 &      1.051 &      0.995   \\
    (28,4287)&      0.762 &      0.333 &      0.475 &      0.999 &      1.032 &      0.954   \\
    (29,4006) &      0.822 &      0.315 &      0.501 &       0.997 &      0.968 &      1.025 \\
\hline
   (201, 57) &      0.873 &      0.254 &      0.606 &      0.992 &      1.027 &      0.985   \\
  (208, 44) &      0.857 &      0.309 &      0.543 &      0.976 &      1.024 &      0.975   \\
  (222, 18) &      0.642 &      0.258 &      0.663 &      0.969 &      1.011 &      0.974   \\
  (252, 23) &      0.682 &      0.269 &      0.597 &      0.985 &      1.027 &      0.979   \\
  (259, 34) &      0.731 &      0.285 &      0.656 &      0.980 &      1.026 &      0.976   \\
  (297, 56) &      0.784 &      0.317 &      0.518 &      0.997 &      1.057 &      0.990   \\
  \hline
  \end{tabular}}
\ec
\end{table}

We then evaluate the relative performance of BLB  and SDB under the same computational budget.   Here, the original SDB corresponds to the cases with $n=\lfloor N^{0.7}\rfloor$ and $R$ is further calculated based on the time consumption of BLB.
We train the optimal hyperparameter the same way with Section \ref{simu:lr}.
  Towards this,
we  randomly generate  6 different combinations of ($R, B$) values    from   $ \lfloor U(15,30)\rfloor \times \lfloor U(2500,5000)\rfloor $  and another  6 pairs of ($R, B$) from $ \lfloor U(150,300)\rfloor \times \lfloor U(1,80)\rfloor $,  as seen in Table \ref{tab:t7}.   Their optimal specifications are obtained via   \eqref{eq:sb}, \eqref{eq:sdb} and \eqref{eq:blb}  under the multivariate case.
From this table,
we can see clearly that the running times of the optimal BLB and SDB are all similar to those of the original setting with prespecified hyperparameters. On the other hand, both methods with optimally chosen hyperparameters outperform their original settings. Moreover, optimal BLB always performs better than optimal (and also original)  SDB and the average improve margin is 44.9\%.   It is reasonable that optimizing BLB over both parameters is expected to yield better result than just
fixing one of the parameters.

\subsection{ Real  Data Analysis}

To demonstrate the application of our method in data analysis, we consider a US Airline Dataset publicly  available at http://stat-computing.org.
This dataset contains detailed flight information. For this analysis, we take the data in the year of 2008 with a sample size $N=492809$,
and study  how the flight  Distance and  ArrDelay
(arrival delay) affect the  ActualElapsedTime (actual elapsed time) via a linear model.  We preprocess the two covariates by taking the signed-log-transformation $\log|x| \cdot \mbox{sign}(x)$ to eliminate the influence of outliers. We evaluate the performance using a procedure similar to Section \ref{simu:lr}, with the critical difference being  that the variance of the ordinary least-squares   estimator has to be estimated. Towards this, we make use of traditional bootstrap by estimating it as $ {M_0}^{-1} \sum_{m=1}^{M_0}   (\hat \beta^{(m)}- \hat \beta) (\hat \beta^{(m)}- \hat \beta) ^\top$, where $\hat \beta^{(m)}$ is the estimate obtained on the $m$ th sample done  by sampling with replacement. The number of bootstrap resamples is taken as $M_0=10000$.

We set $n=\lfloor N^{0.7}\rfloor$, and generate 5 different $(R,B)$ combinations
   from   $ \lfloor U(15,30)\rfloor \times \lfloor U(2500,5000)\rfloor $  and another  5 pairs of ($R, B$) from $ \lfloor U(250,500)\rfloor \times \lfloor U(1,80)\rfloor $,  as seen in Table \ref{tab:realdata}.
   We repeat the experiment 50 times under each setting. Again, we use the MSE  defined in Section \ref{simu:lr} as the performance measure.  From Table \ref{tab:realdata},
we can see clearly that the running times of the optimal BLB and SDB are all similar to those under the original settings with prespecified hyperparameters. On the other hand, BLB and SDB with hyperparameters tuned by our approach outperform their original settings.  Moreover,  the optimal BLB outperforms optimal SDB (and  also SB, results not shown)   under the same time cost.

\begin{table}[!h]
\bc\emph{}
\caption{\label{tab:realdata} Detailed  results  for the  Airline Data. The explanation
of the column headings can be found in the caption of Table \ref{tab:t6}. }
\vspace{0.1cm}
{\small
\begin{tabular}{c|ccc|ccc}
\hline
\hline
 \multicolumn{1}{c}{Original} &  \multicolumn{3}{c}{MSE Ratios}  &  \multicolumn{3}{c}{CPU Time Ratios}\\
  ($R$, $B$)& $\mbox{BLB}^*$/BLB&
   $\mbox{SDB}^*$/SDB & $\mbox{BLB}^*$/$\mbox{SDB}^*$ &$\mbox{BLB}^*$/BLB& $\mbox{SDB}^*$/SDB&  $\mbox{BLB}^*$/$\mbox{SDB}^* $ \\
\hline
(20,3559)&  0.806 &      0.697 &      0.509 &     0.996 &      0.994 &      0.967   \\
(21,4287)&  0.840 &      0.723 &      0.582 &     0.998 &      0.993 &      0.992   \\
(24,3872)&  0.881 &      0.689 &      0.576 &     1.005 &      0.998 &      1.003   \\
(28,4006)&  0.883 &      0.686 &      0.630 &     1.001 &      1.000 &      0.988   \\
(29,3862)&  0.890 &      0.736 &      0.613 &     1.018 &      0.953 &      0.907 \\
  \hline
(398,74) &      0.926 &      0.611 &      0.479 &       0.977 &      0.979 &      0.966   \\
(413,45) &      0.911 &      0.670 &      0.498 &       0.941 &      0.945 &      0.927   \\
(413,15) &      0.785 &      0.635 &      0.440 &       0.920 &      0.933 &      0.921 \\
(437,17) &      0.781 &      0.696 &      0.488 &       0.922 &      0.923 &      0.924   \\
(491,25) &      0.848 &      0.681 &      0.470 &       0.928 &      0.922 &      0.936   \\
  \hline
  \end{tabular}}
\ec
\end{table}

\section{ Conclusion}

In this article we propose a hyperparameter selection approach that can be applied to subsampling bootstrap methods.  A novelty of the approach is to formulate the problem of finding optimal hyperparameters as an optimization one, for which closed-form solutions can be readily obtained using simple, intuitive, almost back-of-envelope calculations. Our extensive simulation study confirms that our hyperparameter selection approach improves efficiency of an estimator without increasing the computational burden. We emphasize, as we have discussed in the Introduction, that bootstrap methods are preferred when analytical formula are unavailable or not easily computable. In view of this, the examples examined in this paper should be taken as proof of concept.

This paper focuses on those estimators whose computational cost is linear in the number of distinct observations in a dataset. It is straightforward to extend our methodology to estimators whose computational cost is of the order $O(n^\gamma)$ for some $\gamma\ge 1$ based on a dataset of size $n$. This setup includes all the computationally feasible estimators in statistics that can be computed in polynomial time. To see how the extension can be done, for the SB and SDB, we just need to solve:
 $ \arg\min_{n, R}~\{c_1'/R+c_2'/n^2\}$  under the constraint  $\alpha   n^\gamma  R \le C_{\max},
 $
 where the constraint is on $n^\gamma R$ instead of $nR$ as in the estimators discussed in Section \ref{sec: optimal}. Here,  $\alpha$ is either  $\alpha_{\mbox{\tiny SB}}$ for SB or $\alpha_{\mbox{\tiny SDB}}$ for SDB, and $c_1'$ and $c_2'$ are estimator specific  parameters.
 This formulation yields the same optimal hyperparameters as in \eqref{eq:sb}.
 Likewise, for the BLB, we replace $nRB$ with \eqref{blb:opt}
 by $n^\gamma RB$. Derivations similar to \eqref{blb:opt} provide the optimal $(B, R)$ as
 $ R^*= \lfloor C_{\max}/(\alpha_1 n^\gamma B^* +\alpha_2 n)\rfloor, \quad B^*=\lfloor  (\wt c_1 \alpha_2 /\wt c_2\alpha_1)^{1/2}  n^{1-\gamma/2}\rfloor,$
  where $\wt c_1$ and $\wt c_2$ are estimator specific parameters, and $\alpha_1$ and $\alpha_2$ are coefficients determining the computational cost as in  \eqref{blb:opt}.

We now discuss an extension of our approach to quantile regression. Specifically,  we examine linear quantile  regression  for illustration by considering   the  model:
$Y_i=X_i\beta_{\tau}+\ve_i$,
where $\beta_{\tau}$  is  a $d$-dimensional coefficient vector, $\ve_i\in \mR^1$ is the random noise such that
$P(\ve_i\leq 0|X_i)=\tau$
 and $\tau\in (0,1)$. That is,  $X_i\beta_{\tau}$ is the $\tau$ th quantile of $Y_i$ given $X_i$. Using the quantile regression approaches in  \cite{koenker2005}, \cite{Wang2007}, and \cite{Chen2019}, we know that $\hat \beta_{\tau} $, the quantile regression estimator, admits the following asymptotic expression:
\beqr
\hat \beta_{\tau} =  \beta_{\tau} + \frac{1}{ f(0)} \Big(\sum_{i=1}^N  X_i X_i^\top \Big)^{-1}
\Big[ \sum_{i=1}^N  X_i \big\{\tau- I(\ve_i \leq 0)  \big\}  \Big]+o_p( N^{-1/2}). \label{eq:1}
\eeqr
Intuitively, if the smaller-order term in \eqref{eq:1} can be ignored, the quantile regression estimator can be expressed as $g(\mu_1,\mu_2)$, a smooth function of two moment estimators $\mu_1 =\sum_{i=1}^N  X_i X_i^\top$ and
$\mu_2 =\sum_{i=1}^N  X_i \{\tau- I(\ve_i \leq 0) \} $. Thereafter,  our theory in Section \ref{sec:osb} can be potentially applied, though justifying it is challenging due to  the non-smoothness of the loss function in quantile regression. We leave it for a future research topic.

Finally, we find the optimal estimator in terms of its MSE in this paper, but it need not be the only choice. Our framework can be extended to other efficiency measures by quantifying the relationship between the choice of efficiency measure and the hyperparameters and then formulating an optimization problem in a manner similar to that used in this paper.

\section*{Acknowledgments}
We are grateful to the Co-editor, Christian Hansen, AE  and two referees for their insightful comments and suggestions.
Yingying Ma's research is partially supported by National
Natural Science Foundation of China (No.12171020, 11801022).
Chenlei Leng's research is partially supported by EPSRC (EP/X009505/1).
Hansheng Wang's research is partially supported by National Natural Science Foundation of China (12271012, 11831008) and also  partially supported by the Open Research Fund of Key Laboratory of Advanced Theory and Application in Statistics and Data Science (KLATASDS-MOE-ECNU-KLATASDS2101).
\section*{Supplementary Material}

\noindent
 Appendix A contains detailed  proofs for Theorem 1--8 and technical details for general parameters and statistics.
 Appendix B contains additional simulation results. Appendix C extends  the optimal hyperparameter selection for bootstrap under  distributed systems.
\singlespacing




\end{document}